\documentclass[iicol,sn-basic]{sn-jnl}

\jyear{2022}%

\theoremstyle{thmstyleone}%
\newtheorem{theorem}{Theorem}
\theoremstyle{thmstyletwo}%

\theoremstyle{thmstylethree}%

\usepackage{graphicx}
\usepackage{enumerate}
\usepackage{natbib}
\usepackage{url}
\usepackage{color}
\usepackage{ulem}
\usepackage{bm}
\usepackage{multicol}
\usepackage{blindtext}

\usepackage[T1]{fontenc} 
\usepackage[utf8]{inputenc} 
\usepackage[english]{babel}
\usepackage{amsmath,amssymb}
\usepackage{mathtools}

\usepackage{microtype} 
\usepackage{rotating} 
\usepackage{booktabs}  
\usepackage{array}     
\usepackage{multirow} 
\usepackage{enumitem}

\newcommand{\TT}{\mathbf{T}}

\newcommand{\Bbeta}{{\boldsymbol{\beta}}}

\newcommand{\RR}{{\rho}}

\begin{document}

\title[False discovery rate envelopes]{False discovery rate envelopes}


\author*[1]{\fnm{Tom\' a\v s} \sur{Mrkvi\v cka}}\email{mrkvicka.toma@gmail.com}

\author[2]{\fnm{Mari} \sur{Myllym{\"a}ki}}\email{mari.myllymaki@luke.fi}
\equalcont{These authors contributed equally to this work.}

\affil*[1]{\orgdiv{Faculty of Economics}, \orgname{University of South Bohemia}, \orgaddress{\street{Studentsk\'a 13}, \city{\v{C}esk\'e Bud\v{e}jovice}, \postcode{37001}, \country{Czech Republic}}}

\affil[2]{ \orgname{Natural Resources Institute Finland (Luke)}, \orgaddress{\street{Latokartanonkaari 9}, \city{Helsinky}, \postcode{FI-00790}, \country{Finland}}}


\abstract{False discovery rate (FDR) is a common way to control the number of false discoveries in multiple testing. There are a number of approaches available for controlling FDR. 
However, for functional test statistics, which are discretized into $m$ highly correlated hypotheses, 
the methods must account for changes in distribution across the functional domain and correlation structure. Further, it is of great practical importance to visualize the test statistic together with its rejection or acceptance region.
Therefore, the aim of this paper is to find, based on resampling principles, a graphical envelope that controls FDR and detects the outcomes of all individual hypotheses by a simple rule: the hypothesis is rejected if and only if the empirical test statistic is outside of the envelope. Such an envelope offers a straightforward interpretation of the test results, similarly as the recently developed global envelope testing which controls the family-wise error rate. Two different adaptive single threshold procedures are developed to fulfill this aim. Their performance is studied in an extensive simulation study. The new methods are illustrated by three real data examples. 
}

\keywords{functional linear model, global envelope test, local spatial correlation, multiple testing, resampling}



\maketitle

\section{Introduction}\label{sec:intro}

\subsection{Motivation and overview}
Nowadays, functional test statistics appear in many applications, for example, in spatial statistics, functional regression, and neuroimaging. A wealth of literature can be found for multiple comparison testing, in particular for micro-arrays applications. However, new challenges arise with functional applications:
(i) the functional test statistics are highly correlated over the functional domain, 
(ii) the distribution of test statistic can change across the domain and even
(iii) the correlation of the test statistic can change across the functional domain.
Further, the distribution of the functional test statistic is rarely known; therefore, nonparametric methods are often required. 
With the functional domain, also the visualization of the test statistic, together with its rejection or acceptance region, is of great practical importance. 

Recently, we developed a global envelope test \citep{MyllymakiEtal2017} which provides a global envelope, i.e., the acceptance region for testing with a functional test statistic, under the control of family-wise error rate (FWER). 
This method relies on resampling to obtain the distribution of the test statistic under the null hypothesis 
in the presence of the high correlation of the test statistics across the functional domain. 
Using rank based ordering of functional statistics as the basis, 
the method handles the changes of the distribution and correlation across the functional domain. 
Further, the acceptance region is visualized in the space of the functional test statistics: 
the parts of the functional domain where the observed test statistics lies outside of the envelope show the reasons for the rejection of the tested global null hypothesis. 
This direct visualization helps the user to interpret the results in more details, especially because resampling allows to use of any test statistics and not just the conventional standardized test statistics. Consequently, this method allows for a very general application while controlling FWER. Our aim in this paper is to define a procedure that does the same job while controlling the false discovery rate (FDR).
To the best of our knowledge, none of the available FDR methods provide such direct visualization with the test result.

Indeed, in functional tests, it is often essential not only to know if the global test is significant but also to estimate the whole domain which is responsible for the rejection, i.e., the local inference. 
A very popular and powerful control for local inference is the FDR. 
An attempt to introduce the FDR control to the infinite-dimensional case was made by \citet{OlsenEtAl2021} as an extension of the FDR procedure of \citet{BenjaminiHochberg1995}. 
However, this procedure is a step-up procedure, whereby it has no direct visualization.
Further, \citet{BenjaminiHochberg1995} method was overcome by newer FDR procedures. 

For the purpose of estimating all alternative hypotheses, the FDR control was defined first in \citet{BenjaminiHochberg1995}. 
More specifically, the difference between FDR and FWER can be seen from Table \ref{tab:def} of outcomes when testing $m$ hypotheses. The FWER is defined to be $\mathbb P(V\geq 1)$, whereas FDR is defined to be $\mathbb E(Q)$, where $Q =\frac{V}{\RR}$ whenever $\RR>0$, and $Q=0$ if $\RR=0$. Clearly, FWER provides more strict control than FDR. Only in the case of all hypothesis being true null (i.e.\ $m_0=m$), the FDR control implies FWER control \citep{BenjaminiHochberg1995}.

\begin{table}
\caption{\label{tab:def}Possible outcomes from $m$ hypothesis tests}
\centering
\begin{tabular}{cccc} \hline
& Accept & Reject & Total  \\ \hline
True null hypothesis & $U$ & $V$ & $m_0$ \\ 
False null hypothesis & $T$ & $S$ & $m_1$ \\ 
Total & $W$ & $\RR$ & $m$ \\ \hline
\end{tabular}
\end{table}

Since \citet{BenjaminiHochberg1995} many theoretical achievements have been made for FDR. \citet{BenjaminiHochberg1995} proved that their linear step-up procedure controls the FDR for independent test statistics. The subsequent article of \citet{BenjaminiYekutieli2001} established the control of FDR for test statistics with more general dependence structures, such as positive regression dependence. \citet{BenjaminiHochberg2000} and \citet{BenjaminiEtAl2006} defined adaptive step-up procedures, which include the estimation of the number of true null hypotheses. \citet{Storey2002} defined another adaptive one-step procedure for controlling FDR.

Since our aim is to define a very general FDR procedure, we stick to the resampling-based methods. 
The first resampling based procedure was defined in \citet{YekutieliBenjamini1999},
where the FDR conditional on $S=s$ (see Table \ref{tab:def}) was estimated. This was criticized by \citet{StoreyTibshirani2001}, and they proposed a resampling method for estimating FDR based on the estimation of the number of true null hypotheses for a single threshold rejection rule. \citet{GeEtAl2003} then rephrased this procedure. \citet{DuboitEtAl2008} defined a Bayes posterior resampling FDR control under the assumption of a mixture model, which assumes that the distributions of all true null statistics are equal and the distributions of all false null statistics are equal. \citet{RomanoEtAl2008} constructed a bootstrap procedure to estimate critical values in a step-down FDR procedure; this method requires bootstrapping the whole data set, which makes its applicability limited. Finally, 
\citet{GeEtAl2008} proposed a resampling procedure based on the step-down procedure of \citet{WestfallYoung1993}, which requires the subset pivotality and relies on the assumption that the joint distribution of statistics from the true nulls is independent of the joint distribution of statistics from the false nulls. Since our goal is direct visualization and step-down procedures cannot be used for this purpose, the methods of \citet{YekutieliBenjamini1999} and \citet{StoreyTibshirani2001} can be adapted to our aim.   

The following software can be used for the computation of FDR: The R/Bioconductor package multtest \citep{GilbertEtal2009} contains the procedures of \citet{BenjaminiEtAl2006} and \citet{DuboitEtAl2008}. The package fdrtool \citep{Strimmer2008} offers the unified approach for the computation of FDR by a semiparametric method allowing for parametric modeling of tests interdependencies. The FDR has further been modeled by many parametric approaches, see \citet{Strimmer2008} and the references therein. The R/Bioconductor package qvalue \citep{qvalue} determines the $q$-value of \citet{Storey2002} for given pointwise $p$-values. The $q$-value gives the scientist a multiple hypothesis testing measure for each observed statistic with respect to FDR. This corresponds to the multiple testing adjusted $p$-value in the FWER sense.
Implementation of the FDR envelopes proposed in this work is provided in the development version of the R library GET \citep{MyllymakiMrkvicka2020} available at \url{https://github.com/myllym/GET}.

\subsection{A comparison of FWER and FDR envelopes: a motivational example}\label{sec:motivation}

\begin{figure}
    \centering
    \includegraphics[width=0.5\textwidth]{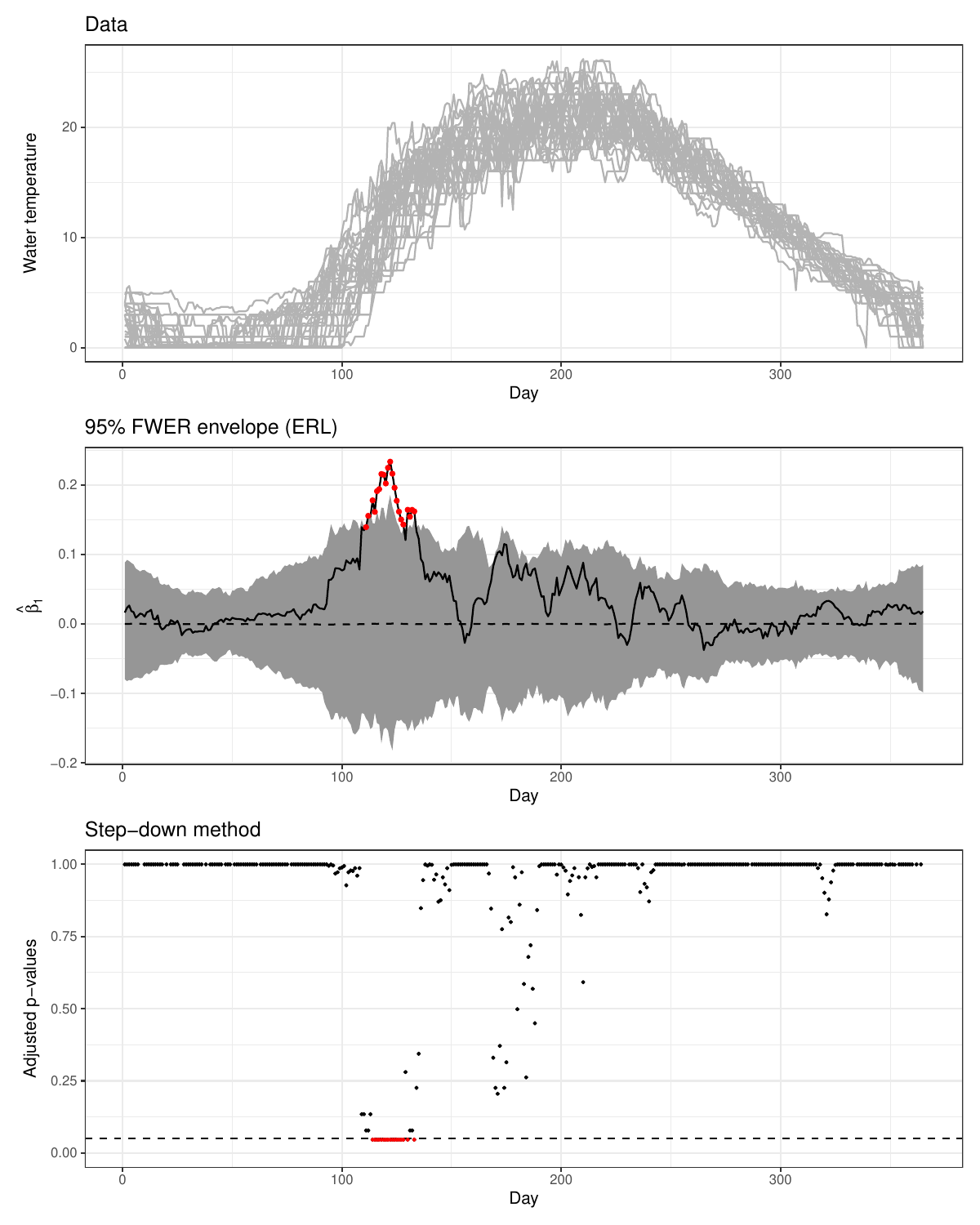}
    \caption{\label{fig:rimov}Annual water temperature curves sampled at the water level of Rimov reservoir in Czech republic every day from 1979 to 2014 (top), the output of the global ERL envelope test ($p<0.001$) for testing the effect of the year on the temperatures (middle), and the step-down $p$-values of \citet{WestfallYoung1993} (bottom). In the middle figure, the grey area represents the 95\% global ERL envelope. The red dots show the days where the data function (solid line) exceeds the envelope, and the dashed line is the mean of simulated functions. At the bottom, dots are $p$-values (red if $\leq 0.05$) and the dashed line represents the level $\alpha=0.05$.}
\end{figure}

To illustrate the merits of the FDR global envelopes, we explore linear trends in annual water temperature curves sampled at the water level of Rimov reservoir in the Czech republic every day from 1979 to 2014 (Figure \ref{fig:rimov}, top). 
Viewing the observations as daily samples of functional data, we fit the model
\begin{equation}\label{eq:rimovmodel}
y_i(k) = \beta_0(k) + \beta_1(k) (i-1978) + e(k),
\end{equation}
where $y_i(k)$ is the water temperature at day $k$, $k=1, \dots, 365$, in year $i$, $i=1979, \dots, 2014$, $\beta_0(k)$ and $\beta_1(k)$ are model parameters, and $e(k)$ denotes the error. To test the null hypothesis 
\begin{equation}\label{eq:rimovH0}
\beta_1(k) = 0 \quad\text{for all } k = 1, \dots, 365,
\end{equation}
we use the estimator of the regression coefficient, i.e.\ $\hat{\beta}_1(k)$, as our test statistic following the methodology proposed in \citet{MrkvickaEtal2021a}.
Then 7000 resamples of the test statistic were obtained under the null hypothesis using simple permutation of raw functional data. We utilized the R packages GET \citep{MyllymakiMrkvicka2020} and pppvalue \citep{pppvalue, XuReiss2020} to compute the three different tests explained below.

Figure \ref{fig:rimov} (middle) shows the output of the global extreme rank length (ERL) envelope test \citep{MyllymakiEtal2017, NarisettyNair2016, MrkvickaEtal2020} on testing the global null hypothesis \eqref{eq:rimovH0} under the FWER control. The global ERL test shows the behavior of $\hat{\beta}_1(k)$ estimated from the data in comparison to the 95\% global envelope constructed from the permutations and detects a significant increase in temperature for 24 days around the 120th day of the year (21 days).  
So it directly identifies the reason for the rejection (increase), and further, it expresses the evolution of the variability of the test statistic under the null hypothesis via the global envelope. It also specifies the amount of the increase over the threshold given by the global envelope. However, the global envelope is not designed for local inference. 
Therefore, the step-down method of \citet{WestfallYoung1993} can be more powerful, but only if many permutations are used. 
The step-down $p$-values \citep{XuReiss2020} are shown in Figure \ref{fig:rimov} (bottom). Using 7000 permutations, we could obtain significant increases (or decreases) for 17 days. 
Repeating the experiment for 2000 permutations, the results of the FWER (ERL) and FDR envelopes produced rather the same results, but none of the FWER adjusted step-down $p$-values were below 0.05.
Thus, the step-down procedure requires plenty of permutations, and it provides only visualization of $p$-values.

\begin{figure}
    \centering
    \includegraphics[width=0.5\textwidth]{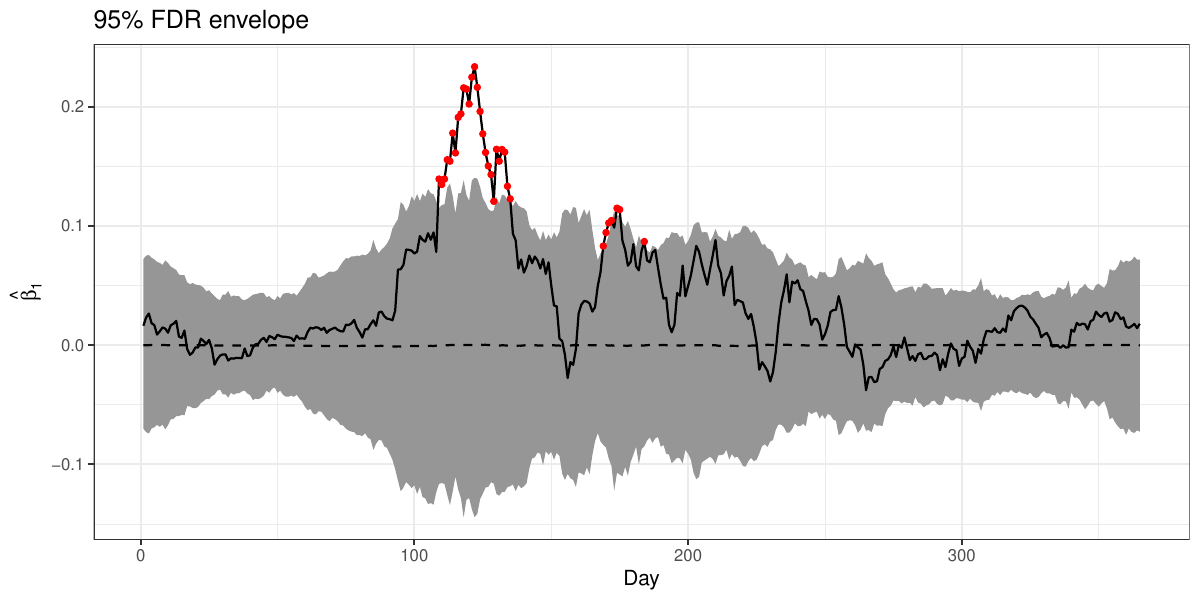}
\caption{\label{fig:rimov_fdr}The 95\% FDR envelope (grey area; obtained by the IATSE algorithm) for testing the effect of year on water temperature in the data presented in Figure \ref{fig:rimov}. The red points denote the days with significant increase in the temperature.}
\end{figure}

Both the global ERL test as well as the step-down method control FWER. 
The aim of this paper is to employ the FDR control instead, retain the direct visualization (single-step methods) and not require plenty of permutations. For example, in applications in spatial statistics or neuroimaging, the simulations of the null model can be rather complex computationally or in terms of memory.
Figure \ref{fig:rimov_fdr} shows an FDR envelope proposed in this work (IATSE algorithm presented in Section \ref{alg:IATSE}).
The test controlling FDR rejected null hypotheses in springtime, around the 120th day of the year as the FWER envelope test, but additionally also in summertime around the 180th day of the year. The null hypothesis was rejected for 35 days in total under FDR control.

\subsection{Outline of the work}

Because the data are assumed to be highly dependent, and the aim is to use the envelope for any functional test statistic, nonparametric resampling methods will be used.
We also assume a general model for alternative hypotheses, since, e.g., in functional permutation tests, the differences between groups of functions will range from zero (null hypothesis) to small (alternative hypothesis) and large ones (alternative hypothesis). Since our aim is to develop an FDR envelope that allows for direct visualization, we will concentrate on a single threshold multiple testing procedure, such as developed by \citet{Storey2002} or \citet{YekutieliBenjamini1999}.

We will propose two new two-stage procedures, with the basis in the algorithm of \citet{Storey2002} to estimate FDR for a given rejection region. 
Essential new features of our algorithms are the following: 
1) an estimate of the probability that a hypothesis is a true null which works well for highly dependent hypotheses, and 2) defining the rejection regions, for which the FDR is estimated, using a new min-max envelope. 
As such we will obtain an envelope which 
(a) allows for direct visualization, 
(b) is directly able to identify all the false hypotheses under the control of FDR, 
(c) is not sensitive to the change of the distribution of the test statistic across the functional domain, and 
(d) is not sensitive to the change of the correlation structure across the functional domain.

The setup for multiple hypothesis testing and the proposed rejection (and acceptance) regions are described in Section \ref{sec:multipleH} and \ref{sec:envelope}.
Section \ref{sec:FDRenvelopes} proposes two algorithms for the FDR envelopes together with $q$-value definition for them. An extension of this method for the case where nuisance parameters have to be estimated is introduced in Appendix \ref{sec:composite}. Section \ref{sec:SS} contains a thorough simulation study comparing the performance of proposed methods with several FDR alternatives.  
The advantage of direct visualization is emphasized by a data example in Section \ref{sec:population}. An example from spatial statistics with a two-dimensional test statistic studying local spatial correlation is presented in Section \ref{sec:localcor}. Conclusions are given in Section \ref{sec:discussion}.


\section{Multiple hypothesis testing}\label{sec:multipleH}

Assume the functional hypothesis testing problem, with hypotheses $H_x$, $x \in \mathbb R^d$, which is discretized into $m$ hypotheses. 
Assume now that all $H_x$ are simple hypotheses. The discretization can be arbitrary; equidistances are not required. 
Thus we assume that we are testing $m$ hypotheses $H_k$, $k=1, \ldots , m$, simultaneously. Denote $H_k=0$ if the $k$-th hypothesis is a true null hypothesis and $H_k=1$ if the $k$-th hypothesis is a false null hypothesis. The sets $\mathcal M_0=\{k:H_k=0\}$ and $\mathcal M_1=\{k:H_k=1\}$ are unknown and the aim is to estimate them. Let us denote by $m_0=\mid\mathcal M_0\mid$ and $m_1=\mid\mathcal M_1\mid$ the numbers of true and false hypotheses. The number of rejected hypotheses $\RR$ is observed together with the number of not rejected hypotheses $W$. On the other hand, the numbers of rejected and not rejected true null hypotheses $V$ and $U$ and the numbers of rejected and not rejected false null hypotheses $S$ and $T$ are not observed (see Table \ref{tab:def}).   

Every single test $k$ is tested by the statistic $T_k$. The vector of the statistics, $\TT=(T_1,\dots,T_m)$, can have an arbitrary distribution. The marginal distributions of $T_k$, $k=1,\dots,m$, can be interdependent and the distribution of $T_k$ can change across $k$. Also the interdependence can vary across $k$.  
A usual approach is to consider the parametric statistics as $t,F,\ldots $, but since such a parametric statistic can change its marginal distribution across $k$, we will switch to rank statistics of these parametric statistics. Obviously, when the statistic changes its marginal distribution, the simultaneous tests will be blind to alternatives that occur for $k$ with smaller variability of test statistic $\TT$; the rank statistics overcome this problem.

When $\mathcal R$ denotes the rejection region in a single test, 
$\mathbb P(T \in \mathcal R \mid H=0) $ is called the type I error rate and is controlled by a predetermined level $\alpha$. On the other hand, $\mathbb P(T \notin \mathcal R \mid H=1)$ is called the type II error rate and is not controlled. 

In multiple testing, the problem of controlling the type I error is more complex, and it is generalized in several ways. The most common is FWER, where the probability of at least one type I error is controlled, i.e., \ FWER$=\mathbb P(V\geq 1)$. But also, per comparison error rate can be defined as PCER$=\mathbb E(V)/m$ or per family error rate PFER$=\mathbb E(V)$. In functional setting, \citet{PiniVantini2017} recently introduced the interval-wise error rate in order to control the type I error for each subinterval of the functional domain. 
All the above-mentioned controls are designated to control the existence of rejection of a true null hypothesis, whereas the famous FDR was introduced in order to determine $\mathcal M_0$ and $\mathcal M_1$. The natural way to control FDR would be by $\mathbb E(V/\RR)$, but the case of division by zero has to be treated. Therefore, the first definition of FDR \citep{BenjaminiHochberg1995} was FDR$=\mathbb E(\frac{V}{\RR} \mid \RR>0)\mathbb P(\RR>0)$. Then \citet{Storey2002} defined positive false discovery rate by pFDR$=\mathbb E(\frac{V}{\RR}\mid\RR>0)$, which controls error rate when positive findings occurred.

The rejection region for the multiple hypothesis test can be given by a step-up procedure, subsequently rejecting the hypothesis from the smallest $p$-values, as proposed by \citet{BenjaminiHochberg1995}. Such rejection region is estimated to fulfill the FDR control. 
If an opposite approach is taken, it is possible to estimate FDR or pFDR for a general fixed rejection region $\Gamma=(\mathcal R_1, \ldots , \mathcal R_m)$. \citet{Storey2002} proposed to use a single value threshold $\gamma$ as a natural rejection region for the continuous statistics $T_k$. While we would like to use rank statistics, this approach brings many ties and has to be refined. For example, assume that the null hypothesis is resampled $s$ times. Then the single test statistic can achieve only ranks from 1 to $s$. Consequently, the single threshold rejection region can also obtain only these values, and the corresponding estimated FDR values can be rather sparse, leading to only approximate control.
For this purpose, we will define new acceptance regions based on the global rank envelope \citep{MyllymakiEtal2017}. 

We note that the problem of ties in $p$-values for non-resampling algorithms was also discussed recently in the literature \citep{ChenEtAl2017, Chen2020}. They assumed that the test statistic comes from a given discrete distribution.

\section{Continuous min-max envelope}\label{sec:envelope}

In our earlier works, summarized in \citet{MyllymakiMrkvicka2020}, we have defined several global envelopes, which are appropriate in the case of the FWER control, where it is desired to construct the envelopes together with the data and simulated functions in order to get the exact Monte Carlo test in the Barnard sense \citep{Barnard1963}. 
Here we introduce a new min-max envelope which is a continuous version of the rank envelope proposed in \citet{MyllymakiEtal2017}, and use it as the basis of our FDR envelopes.

We will estimate the FDR for a given rejection region $\Gamma$ which corresponds to the acceptance region determined by the envelope $\mathcal E$. 
We introduce a new, continuously enlarging envelope that is
controlled by a single parameter, similarly as it is in the classical constant rejection rule \citep{Storey2002}. 
However, our continuous min-max envelope reflects the shape of the distribution of the test function under the complete null hypothesis.

Assume now, that we have $s$ resamples of the complete null hypothesis, i.e.\ $m_0=m$, which gives us $s$ test vectors $\TT_i=(T_{i1}, \ldots , T_{im}), i=1, \ldots , s$. 
Let us define the lower and upper bounds of the envelope $\mathcal E_\gamma$ for the two-sided alternative hypotheses.
For $\gamma$ being an integer, $\gamma=1,2,3,\dots, \lfloor s/2 \rfloor$, the bounds are defined by 
$$
\mathcal E_{\gamma,k}^{\text{low}} = \text{min}^\gamma \{T_{ik}: i=1,\ldots,s\}, $$
$$ 
\mathcal E_{\gamma,k}^{\text{upp}} = \text{max}^\gamma \{T_{ik}: i=1,\ldots,s\}
$$
for $k=1,\dots,m$,
where $\text{min}^\gamma$ and $\text{max}^\gamma$ denote the $\gamma$-th smallest and largest values, respectively.
For a non-integer $\gamma > 1$, we define the bounds of $\mathcal E_\gamma$ as the average of neighboring integer $\gamma$s:
$$
\mathcal E_{\gamma,k}^{\text{low}} = \mathcal E_{[\gamma],k}^{\text{low}} + (\gamma-[\gamma])\cdot (\mathcal E_{[\gamma]+1,k}^{\text{low}}-\mathcal E_{[\gamma],k}^{\text{low}}), $$
$$
\mathcal E_{\gamma,k}^{\text{upp}} = \mathcal E_{[\gamma],k}^{\text{upp}} - (\gamma-[\gamma])\cdot (\mathcal E_{[\gamma],k}^{\text{upp}}-\mathcal E_{[\gamma]+1,k}^{\text{upp}}).
$$
For $0<\gamma < 1$, we define the envelope $\mathcal E_\gamma$ as an extension of $\mathcal E_{1}$:
$$
\mathcal E_{\gamma,k}^{\text{low}} = \mathcal E_{1,k}^{\text{low}} + \log \gamma \cdot  (\mathcal E_{1,k}^{\text{upp}}-\mathcal E_{1,k}^{\text{low}}), $$
$$
\mathcal E_{\gamma,k}^{\text{upp}} = \mathcal E_{1,k}^{\text{upp}} - \log \gamma \cdot  (\mathcal E_{1,k}^{\text{upp}}-\mathcal E_{1,k}^{\text{low}}).
$$

The definition for the one-sided alternatives can be obtained accordingly by omitting the lower or upper part.

\section{FDR estimation}\label{sec:FDRenvelopes}

This section describes algorithms for FDR estimation from $s$ resamples of the complete null hypothesis. We adopt the \citet{Storey2002} approach of estimation FDR, rather than the \citet{BenjaminiHochberg1995} approach of estimation of rejection region. This allows us to estimate FDR for a general global envelope $\mathcal E$, which represents the acceptance region corresponding to the rejection region $\Gamma$.  
The advantage of our proposed algorithms is that they make no assumptions on the correlation structure of the test statistics and also not on the false null hypotheses. The only assumption here is that the true null hypotheses are simple.

For a fixed $\Gamma$, we can observe only $\RR(\Gamma)$, the number of positives, together with the corresponding numbers of acceptances, $W(\Gamma)$. The number of false positives $V(\Gamma)$ is not observable, but it can be estimated.

\citet{StoreyTibshirani2001} showed that 
\begin{equation}\label{StoreyTibshirani_approximation}
\mathbb E \left[ \frac{V(\Gamma)}{\RR(\Gamma)}\right] \approx \frac{\mathbb E V(\Gamma)}{\max (\RR(\Gamma),1)} = \pi_0 \frac{\mathbb E \RR^0(\Gamma)}{\max (\RR(\Gamma),1)},
\end{equation}
where $\RR^0(\Gamma)$ is the number of positives under complete null hypothesis. 
In order to estimate the probability that a hypothesis is true null, $\pi_0$, \citet{StoreyTibshirani2001} proposed to choose a smaller rejection region $\Gamma'$ and compute  
\begin{equation} \label{eq:pi0}
\hat\pi_0 =  \frac{ W(\Gamma')}{\mathbb E W^0(\Gamma')}.
\end{equation}
When working with highly correlated data, this estimator of $\pi_0$ has a significant drawback: the histogram of $p$-values observed under the null hypothesis is not uniform, which is assumed by the method \citep{StoreyTibshirani2003}. This feature causes the estimator to fail for any choice of $\Gamma'$, even for the smoothing method defined in \citet{StoreyTibshirani2003}. Also, \citet{XieEtAl2005} claimed that estimation of $\pi_0$ is a challenging problem. In non-resampling approaches, the choice of $\Gamma'$ attracted many researchers in order to define so-called adaptive FDR procedures based on the Storey estimator \eqref{eq:pi0} \citep[e.g.][]{LiangNettleton2012, HeesenJanssen2016}. However, since these procedures are still based on the basic Storey estimator, we searched for another estimator of $\pi_0$, which would not require the uniformity of $p$-values.
\citet{Hwang2011} compared several estimators from which the lowest slope method developed by \citet{BenjaminiHochberg2000} and the mean of differences method \citep{HsuehEtAl2003} seem to be the best in the dependent case, but \citet{Hwang2011} did not include in the comparison the two-stage approach of \citet{BenjaminiEtAl2006}. 
The two-stage method was compared to the method of \citet{BenjaminiHochberg2000} in \citet{BenjaminiEtAl2006}, and it was found to be significantly better. Also, this two-stage approach satisfies the conservative estimation of $\pi_0$ under the independence of all hypotheses. 
For these reasons, 
we utilize the $\pi_0$ estimation of this two-stage approach of \citet{BenjaminiEtAl2006}, which is defined in the following algorithm, in one of our proposed algorithms, and we also use the algorithm as the reference method in the simulation study.
In our second algorithm, we use another estimator for $\pi_0$, and we prove its strong conservativeness without further assumptions.

\begin{algorithm}\label{alg:ATS}
Adaptive two-stage procedure of \citet{BenjaminiEtAl2006} (ATS):
\begin{enumerate}
    \item Use the linear step-up procedure at level $\alpha^*=\alpha/(1+\alpha)$. Let $r$ be the number of rejected hypotheses. If $r=0$ do not reject any hypothesis and stop; if $r=m$ reject all $m$ hypotheses and stop; otherwise continue.
\item Let $\hat\pi_0=(m-r)/m$.
\item Use the linear step-up procedure with $\alpha'=\alpha/\hat\pi_0$.
\end{enumerate}
\end{algorithm}

The linear step-up procedure uses $m$ $p$-values $\{p_1,\ldots , p_m\}$ corresponding to the $m$ hypotheses and the ordered $p$-values $p_{[1]} \leq p_{[2]} \leq \dots \leq p_{[m]}$. It rejects $k$ hypotheses corresponding to the first $k$ smallest $p$-values, where $k=\max\{i:p_{[i]}\leq i\alpha/m\}$. If such $k$ does not exists, it rejects no hypothesis. 

In the first step of the algorithm, the numbers of rejections are estimated assuming that $\pi_0=1$. In the second step, an estimate for $\pi_0$ is found, and the third step finds the rejections utilizing the estimate of $\pi_0$.

\subsection{Two-stage resampling procedures for FDR envelopes}

We can now, in the two-stage procedure, replace the linear step-up procedure with the procedure of \citet{StoreyTibshirani2001} applied to our rejection regions of Section \ref{sec:envelope}. 
The resulting adaptive two-stage resampling-based procedure (ATSE) defined in the following algorithm has a global envelope representation corresponding to the rejection region $\Gamma_\gamma$.

\begin{algorithm}\label{alg:ATSE}
Adaptive two-stage envelope (ATSE):
\begin{enumerate}
\item Find the largest rejection region $\Gamma_{\gamma^*}$ for which 
$$
\frac{\mathbb E \RR^0(\Gamma_{\gamma^*})}{\max (\RR(\Gamma_{\gamma^*}),1)}\leq \alpha/(1+\alpha).
$$
Let $r$ be the number of rejected hypotheses for $\Gamma_{\gamma^*}$. If $r=0$, do not reject any hypothesis, take $\Gamma_\gamma=\Gamma_{\gamma^*}$ and stop; if $r=m$ reject all $m$ hypotheses, take $\Gamma_\gamma=\Gamma_{\gamma^*}$ and stop; otherwise continue.
\item Let $\hat\pi_0=(m-r)/m$.
\item Find the largest rejection region $\Gamma_\gamma$ for which
$$\frac{\mathbb E \RR^0(\Gamma_\gamma)}{\max (\RR(\Gamma_\gamma),1)}\leq \alpha/\hat\pi_0.$$
\end{enumerate}
\end{algorithm}

As the estimate of $\mathbb E \RR^0(\Gamma_\gamma)$ in the steps 1.\ and 3.\ of the algorithm we use 
\begin{equation}\label{eq:ER0_theo}
    \mathbb E \RR^0(\Gamma_\gamma) \cong m \tilde\gamma, 
\end{equation}
where $\Gamma_\gamma$ is the rejection region corresponding to $\mathcal{E}_{\gamma}$ as defined in Section \ref{sec:envelope} and  $\tilde\gamma$ is the pointwise $p$-value corresponding to $\Gamma_\gamma$. That is,
\begin{equation}\label{gammatilde}
  \tilde\gamma =
    \begin{cases}
      2\gamma/s & \text{in the two-sided case}\\
      \gamma/s & \text{in the one-sided case}.
    \end{cases}       
\end{equation}
Note that there is equality in Equation \eqref{eq:ER0_theo} for an integer $\gamma$. 
A formula similar to \eqref{eq:ER0_theo} was used in \citet{Storey2002} and \citet{StoreyTibshirani2003} for a given threshold $p$.


Instead of using the idea of ATS algorithm to estimate $\pi_0$, it is also possible to use the following Iterative adaptive two-stage envelope (IATSE) algorithm.

\begin{algorithm}\label{alg:IATSE}
Iterative adaptive two-stage envelope (IATSE):
\begin{enumerate}
\item Find the largest rejection region $\Gamma_{\gamma^*}$ for which  
$$
\frac{\mathbb E \RR^0(\Gamma_{\gamma^*})}{\max (\RR(\Gamma_{\gamma^*}),1)}\leq \alpha.
$$ 
Let $r$ 
be the number of rejected hypotheses for $\Gamma_{\gamma^*}$. 
If $r=0$, do not reject any hypothesis, take $\Gamma_{\gamma}=\Gamma_{\gamma^*}$ and stop; otherwise continue. 
\item 
Let $\tilde\gamma^*$ be the pointwise $p$-value corresponding to the $\Gamma_{\gamma^*}$ defined according to the Formula \eqref{gammatilde} and
\begin{eqnarray}\label{iatse_pi0}
\hat\pi_0 = \min\left(1, \frac{(m-r)/m}{1-\tilde\gamma^*}\right). 
\end{eqnarray}
\item Find the largest rejection region $\Gamma_\gamma$ for which 
$$\frac{\mathbb E \RR^0(\Gamma_\gamma)}{\max (\RR(\Gamma_\gamma),1)}\leq \alpha/\hat\pi_0.$$
\end{enumerate}
\end{algorithm}

The expectations in steps 1.\ and 3.\ of the IATSE algorithm are calculated according to Equation \eqref{eq:ER0_theo} similarly as in the ATSE algorithm.
Note here that the ATSE method controls the estimate of $\pi_0$ by assuming a greater significance level in the first step, in the same way as the ATS algorithm of \citet{BenjaminiEtAl2006}. On the other hand, the IATSE method adds the expected number of false null hypotheses to the estimate of $\pi_0$.  

The IATSE algorithm is motivated by the following conservative, iterative estimate of $\hat m_0$: $\hat m_0^1=m-r+m \tilde\gamma^*$ and $\hat m_0^l=m-r+\hat m_0^{l-1}\tilde\gamma^*$. The limit of this iterative procedure is according to the sum of geometrical sequence equal to $(m-r)/(1-\tilde\gamma^*)$.

\begin{theorem}
The estimate \eqref{iatse_pi0} of the IATSE algorithm is a conservative estimate of $\pi_0$.
\end{theorem}

\begin{proof}
It suffices to prove that $\mathbb E \frac{(m-r)/m}{1-\tilde\gamma^*} \geq \pi_0$.
Denote now $\beta_k$ the probability of the error of the second kind for any $k$-th hypothesis which is false. 
\begin{eqnarray*}
 \mathbb E \frac{(m-r)/m}{1-\tilde\gamma^*} &=&  \frac{(m-(m_0\tilde\gamma^*+\sum_{k\in \mathcal M_1}(1-\beta_k)))/m}{1-\tilde\gamma^*}\\
 &=& \frac{(m_0-m_0\tilde\gamma^*+\sum_{k\in \mathcal M_1}(\beta_k))/m}{1-\tilde\gamma^*} \\
 &=& \pi_0 + \frac{\sum_{k\in \mathcal M_1}(\beta_k)/m}{1-\tilde\gamma^*} \geq \pi_0.
\end{eqnarray*}
\end{proof}

Since the strong control of our whole procedures can be studied only under independence assumptions of test statistics \citep[similarly as in][]{LiangNettleton2012}, which is not realistic in functional setting, we resort on checking the control by an extensive simulation study.
Note here that the strong control of the ATS algorithm was also proven for independent hypotheses only \citep{BenjaminiEtAl2006}.

Appendix \ref{sec:composite} contains an algorithm for FDR estimation in case the null hypothesis is not simple, but some nuisance parameters are estimated, which is an extension of the proposed algorithms. Appendix \ref{sec:dep} contains further algorithms and their modifications that incorporate the dependence of the test statistics in the FDR estimation and which can be used for envelope construction. These algorithms were used in the simulation study for comparison with the proposed algorithm, but 
they did not perform as well as the algorithms proposed in this section.

\subsection{$q$-values}

The $q$-value related to the hypothesis $H_k$, $k=1,\dots,m$,
is the size of the test still rejecting $H_k$ while controlling FDR. In the case of FDR envelopes, this size corresponds to the $\tilde\gamma$ value that defines the largest possible envelope (i.e., smallest possible rejection region) such that the hypothesis $H_k$ is rejected. Formally, 
$$
q_k = \inf_\gamma\{ \tilde\gamma:  T_{k} \in \Gamma_\gamma \}.
$$
\citet{YekutieliBenjamini1999} noted that it is advantageous to use adjusted $p$-values in multiple hypothesis testing because then the level of the test does not have to be determined in advance. The same can be said about $q$-values in the context of FDR control.

\section{Simulation study}\label{sec:SS}

A simulation study was performed to investigate the FDR control and power of different FDR methods listed in Table \ref{tab:simstudy_methods}.
We include some methods with parametric and nonparametric pointwise $p$-values to stress differences between them.
We show some methods with known $\pi_0$ in order to check the problem of $\pi_0$ estimation by comparing it to the method with unknown $\pi_0$.
The ATS method is included as the reference method, the best one we have found from the literature. 
ST, YB, YBu, IATSE-YB, IATSE-YBu, ATSE, and IATSE are included in the comparison as candidates for the best FDR envelope. The methods YB, YBu, IATSE-YB, IATSE-YBu, which incorporate the dependence of the test statistics, are described in Appendix \ref{sec:dep}.

\begin{table*}[h]
\begin{center}
\begin{minipage}{\textwidth}
 \caption{\label{tab:simstudy_methods}Multiple testing methods with FDR control investigated in the simulation study}
 \begin{tabular}{ll} \hline
Short name &  Description \\ \hline
ATSp & ATS \citep{BenjaminiEtAl2006} with parametric $p$-values \\
ATS & ATS \citep{BenjaminiEtAl2006} with non-parametric $p$-values \\
STp &  \citet{StoreyTibshirani2003} with parametric $p$-values  \\
ST &  \citet{StoreyTibshirani2003} with non-parametric $p$-values \\
STp/$\pi_0$ & \citet{StoreyTibshirani2003} with parametric $p$-values and known $\pi_0$ \\
ST/$\pi_0$  & \citet{StoreyTibshirani2003} with non-parametric $p$-values and known $\pi_0$ \\
YB  & \citet{YekutieliBenjamini1999} (with $\pi_0=1$) \\
YBu  & \citet{YekutieliBenjamini1999} 'u'-version with $\beta=0.05$ (with $\pi_0=1$) \\ 
IATSE-YB & \citet{YekutieliBenjamini1999} complemented with $\pi_0$ estimation \\
IATSE-YBu  & \citet{YekutieliBenjamini1999} 'u'-version complemented with $\pi_0$ estimation \\
ATSE   & FDR envelope using the ATSE algorithm \\ 
IATSE  & FDR envelope using the IATSE algorithm \\ \hline 
\end{tabular}
\end{minipage}
\end{center}
\end{table*}

The FDR methods were investigated for detecting the difference between two groups of functions, where each group had ten observations. 
The functions in the groups were generated from Gaussian, Student's $t_1$
or bimodal random processes on $[0,1]$ with means $\mu_1(k)$ and $\mu_2(k)$ for the two groups, respectively. 
We considered 11 different mean function combinations for the two groups.
As a null case, we considered the 'no differences' case where both groups had mean zero.
Further, we considered three cases named 'Abrupt', 'Fluctuating' and 'Soft', where the first group had zero mean, while the mean in the second group deviated from zero in different ways and on different sub-intervals of $[0,1]$. 
Finally, a case of 'Bumps' was considered where both mean functions are positive on $[0, 0.66]$, and 0 otherwise. The mean functions slightly differ on $[0, 0.66]$.
More precisely, the mean functions are specified in Table \ref{tab:simstudy_means} and visualized in Figure \ref{fig:simstudy_means} of Appendix \ref{sec:simstudy}. 
The Gaussian process was constructed as $Z_i(k) = \mu_i(k) + e(k)$, where $i=1,2$ and $e(k)$, $k\in[0,1]$, are the Gaussian errors with the standard deviation $\sigma_Z$ and the exponential covariance function with the scale parameter $\phi_Z = 0$ (independent case), $0.1$ or $0.5$.
In the case of Student's $t_1$ distribution, the error was obtained from two independent standard Gaussian errors with the same covariance as $e(k)$.
In the bimodal case, the error was obtained by transforming $e(k)$ as $e_b(k)=\text{sign}(e(k)) \mid\frac{3}{4}e(k)\mid^{1/5}$.
Different values were considered for $\sigma_Z$ as specified in Table \ref{tab:simstudy_means}.
The choices were different for different mean functions in order to reach the powers of the methods in a reasonable range.

\begin{table*}[h]
    \caption{\label{tab:simstudy_means}Mean functions $\mu_1(k)$ and $\mu_2(k)$, $k\in [0,1]$, used in the simulation study. The equations for the bumps are $\mu_1^{\text{bump}}(k) = 3\cdot (5+2)\cdot k \cdot (1-k)^{5+2}$ and $\mu_2^{\text{bump}}(k) = 3\cdot (5+6)\cdot k\cdot (1-k)^{5+6}$ for $k < 0.66$, and 0 otherwise. Different values were considered for the parameters of the mean functions (Parameters) leading to different proportions of the true nulls ($m_0 / m$). The final row gives the standard deviation of the Gaussian error, $\sigma_Z$, used for the Gaussian and bimodal errors.}
\centering
{\small    \begin{tabular}{l|lllll} \hline
Model &    1   & 2-4   & 5-7   & 8-10  & 11\\ \hline
Name  &   No diff. &  Abrupt & Fluctuating &  Soft & Bumps\\
$\mu_1(k)$  &   0  & 0 & 0 & 0 & $\mu_1^{\text{bump}}(k)$ \\
$\mu_2(k)$  &  0 &  $1$ for $s \in (0,a)$,  & $\max(0, \sin(b k)-k)$ & $\exp\{-((k - 0.5)^2/(2 \sigma^2)\}$  & $\mu_2^{\text{bump}}(k)$\\
            &    &  $0$ otherwise          &                        & for $k \in [0.5-3\sigma, 0.5+3\sigma]$, & \\
            &    &                         &                        & $0$ otherwise & \\
Parameters &  &  $a = 0.05$, $0.20, 0.50$ &  $b=20, 50, 100$ & $\sigma = 0.01, 0.025, 0.05$ &  \\
$m_0/m$     &  1 & 0.95, 0.8, 0.5 & 0.65, 0.67, 0.67 & 0.94, 0.85, 0.70  & 0.34 \\
$\sigma_Z$  &  0.5 &  0.5 &  0.3 & 0.3 & 0.05 \\ \hline
    \end{tabular}
    }
\end{table*}

We first compared all the methods of Table \ref{tab:simstudy_methods} with the resolution $m=200$, such that $k$ obtained $m$ equidistant values on $[0,1]$. We did the comparison for different numbers of resamples $s=100,500,1000,2000,5000$, for all 11 models explained in Table \ref{tab:simstudy_means} with the Gaussian, $t_1$ and bimodal errors and with the three strengths of correlation determined by $\phi_Z=0,0.1,0.5$, in total for $3\cdot 3 \cdot 11 \cdot 5 = 495$ cases. 
To test the hypothesis of no difference in the means of the two groups, we used the functional $F$-statistic (the vector of $F$-values calculated for each $k$).
The chosen $F$-statistics allows to compute also parametric $p$-values and compare the performance of parametric and nonparametric FDR methods (see Table \ref{tab:simstudy_methods}). Note that the nonparametric methods are not limited to the $F$-statistic, but other statistics, which can be more easily interpreted, can be used (see Section \ref{sec:population}). 

Each case was repeated 2000 times, and for each repetition, we calculated the FDR, i.e., the number of falsely rejected hypotheses divided by the total number of rejections. Also, the proportion of rejected hypotheses from the known true positives as a measure of power was calculated.
The mean FDR and power as well as their standard errors, $\sigma_{\text{FDR}}$ and $\sigma_{\text{power}}$, were then calculated from the 2000 repetitions.
In the following, we summarize the relevant findings from the simulation study. The rest of the results are given in Appendix \ref{sec:simstudy}.

Figure \ref{fig:res_part1_model14511} shows the mean FDR ($\pm 2\cdot \sigma_{\text{FDR}}$) for Models 1 (complete null model), 4, 5 and 11 with Gaussian errors.
First, the results show that ST and STp methods are highly liberal for correlated cases. This liberality is caused by the $\pi_0$ estimation: the ST methods with known $\pi_0$ are not liberal. 
Some repetition for ST and STp even fell down due to the non-uniform distribution of pointwise $p$-values. Namely, tens of repetitions fell down for correlation 0.1, and hundreds fell down for correlation 0.5; these cases were excluded from the results.
The results also show the liberality of all YB methods for the complete null model (Model 1) with correlated errors. The liberality of YB and YBu methods is then suppressed for Model 5 because the methods do not consider the estimation of $\pi_0$, and this brings conservativeness when $\pi_0<1$.
Adding the estimation of $\pi_0$ into YB methods, i.e., using IATSE-YB and IATSE-YBu methods causes the return to liberality also in cases where $\pi_0<1$, particularly for Models 4 and 11.  
Due to these observed liberalities, we concentrate our further comparison only on ATSp, ATS, ATSE, and IATSE methods, whose estimated mean FDR was satisfactory for all models.

\begin{figure*}[h]
    \centering
\includegraphics[width=\textwidth]{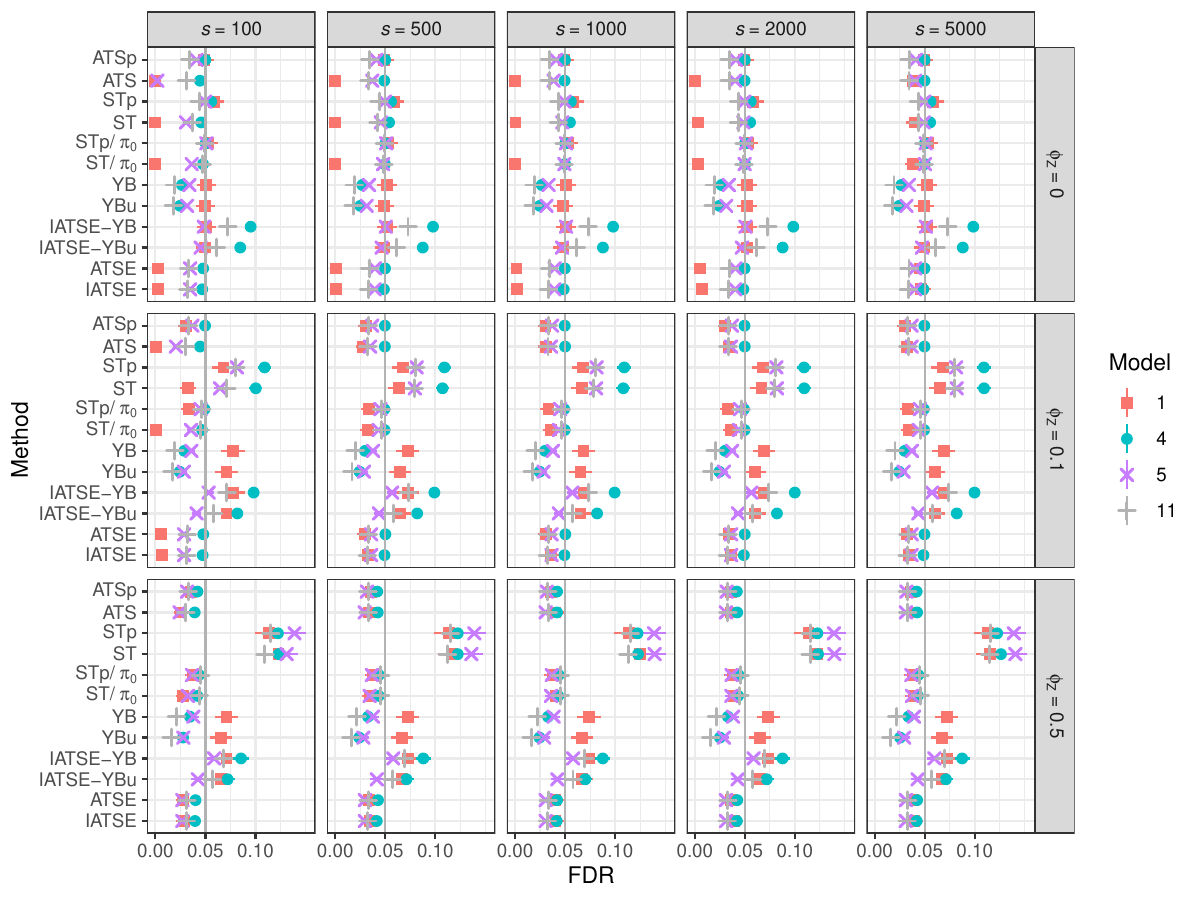}
    \caption{\label{fig:res_part1_model14511}The mean FDR ($\pm 2\cdot \sigma_{\text{FDR}}$ given by bars around dots, often invisible due to being so small) of methods of Table \ref{tab:simstudy_methods} for Model 1, 4, 5 and 11 with Gaussian error and $m=200$ for different numbers of resamples $s$ and correlation parameter $\phi_Z$.}
\end{figure*}

\begin{figure*}[h]
    \centering
    \includegraphics[width=\textwidth]{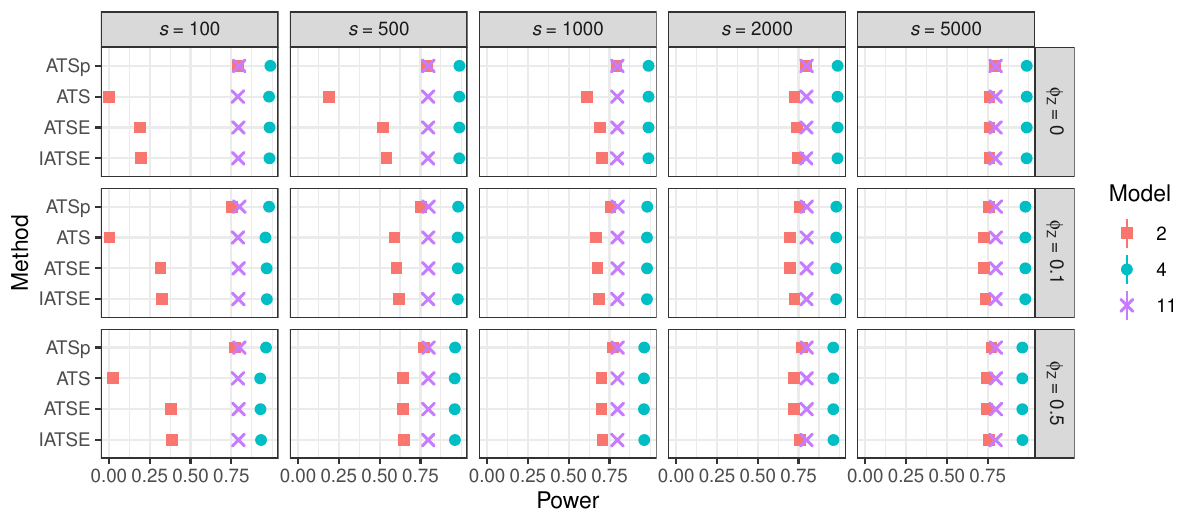}
    \caption{\label{fig:res_part1_power}The mean power ($\pm 2\cdot \sigma_{\text{Power}}$ given by bars around dots) of four methods of Table \ref{tab:simstudy_methods} for Model 2, 4 and 11 with Gaussian error and $m=200$.}
\end{figure*}

The power of non-liberal methods is shown for Models 2, 4, and 11 in Figure \ref{fig:res_part1_power}. Here we can observe, especially for Model 2, that the parametric ATS achieves the highest power, which is only reached by nonparametric methods by applying at least 5000 permutations. The IATSE method seems to be the most powerful from nonparametric methods, which is more apparent for small numbers of permutations $s$.
The number of resamples is not a negligible feature of the test since, for example, in spatial statistics, the resampling of a test statistic can be costly in terms of time. 

\begin{figure*}[h]
    \centering
    \includegraphics[width=\textwidth]{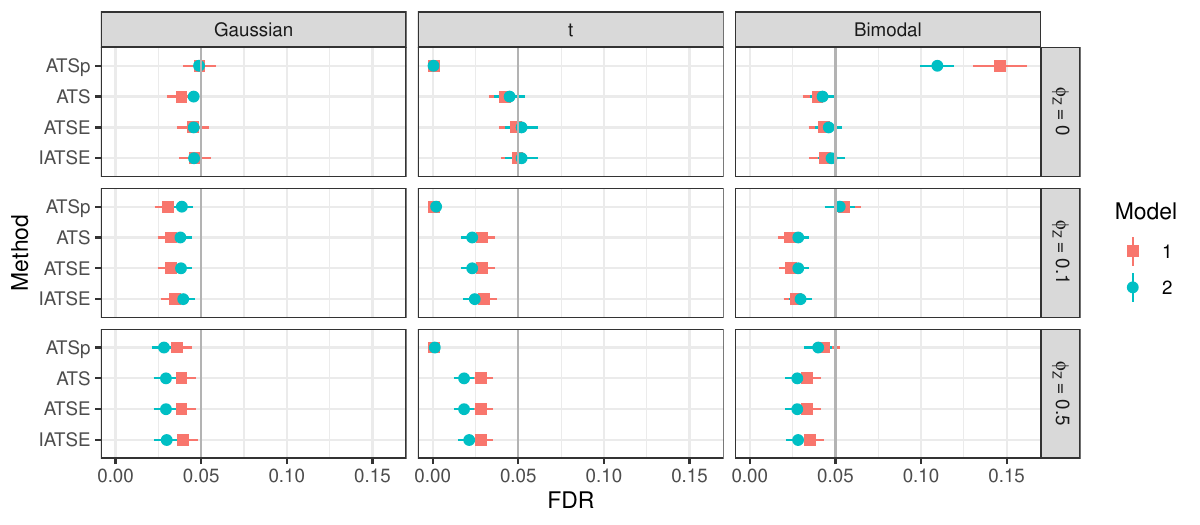}
    \caption{\label{fig:res_errors}The mean FDR ($\pm 2\cdot \sigma_{\text{FDR}}$ given by bars around dots) of four methods of Table \ref{tab:simstudy_methods} for Model 1 and 2 with Gaussian error, heavy tailed error ($t$) and bimodal error for $s=5000$ and $m=200$.}
\end{figure*}

The advantage of the parametric method is clearly lost when the errors are not Gaussian since the parametric $p$-values assume the Gaussianity. Figure \ref{fig:res_errors} shows that if the assumptions of parametric $p$-values are not fulfilled, the parametric method can achieve either strong liberality or strong conservativeness. 

\begin{figure*}[h]
    \centering
    \includegraphics[width=\textwidth]{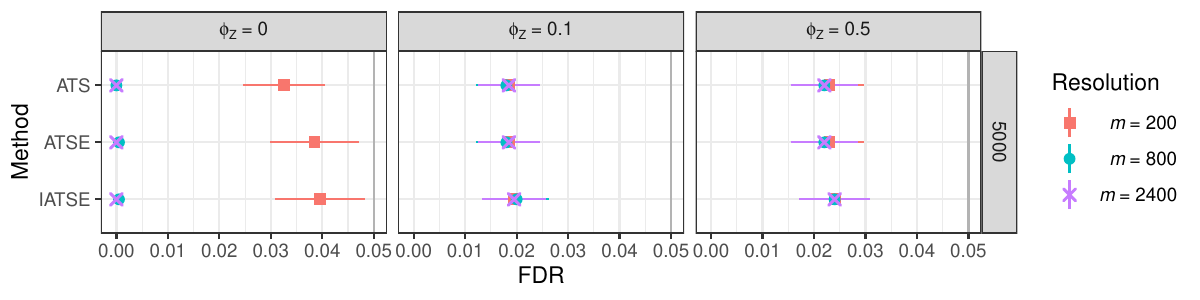}
    \caption{\label{fig:res_part2}The mean FDR ($\pm 2\cdot \sigma_{\text{FDR}}$ given by bars around dots) of four methods of Table \ref{tab:simstudy_methods} for Model 1 with Gaussian error, $s=5000$ and increasing resolution $m$.}
\end{figure*}

The above experiment suggests that the IATSE method is a universal method for FDR control. Namely, it does not require knowing the test statistic distribution, it was conservative in all studied cases, conservativeness of its $\pi_0$ estimation was proven, and it allows for graphical interpretation using the envelopes. Further, it tends to have a little bit tighter control of FDR than ATS and ATSE, as it can be seen from Figure \ref{fig:res_part2} where the FDR control is studied with respect to increasing resolution $m$ for the Gaussian error case. This figure also shows that the increased resolution has no influence on the FDR control of these three methods when the number of resamples is large enough to distinguish the extreme pointwise $p$-values. 

\section{Population growth example}\label{sec:population}

\begin{figure*}[h]
    \centering
    \includegraphics[width=0.8\textwidth]{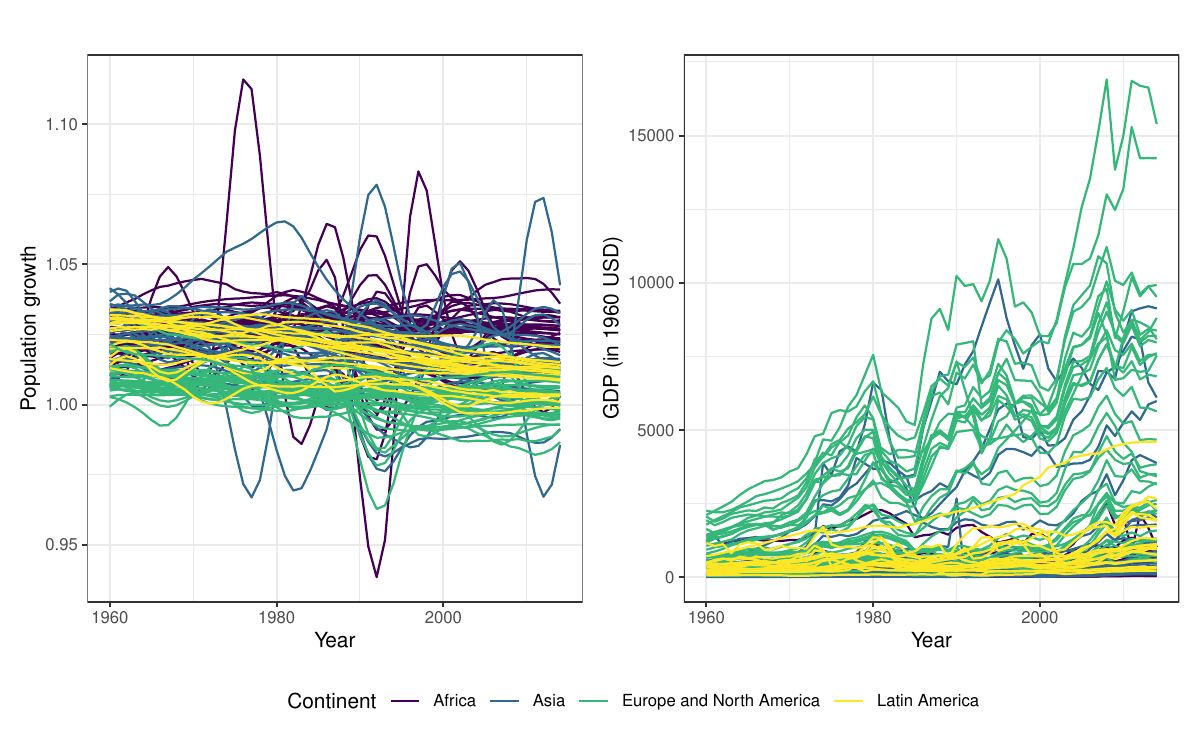}
\caption{\label{fig:popgrowth}Population growth and GDP (in 1960 USD) curves for 1960--2014 for 112 countries from the four continents.}
\end{figure*}

Figure \ref{fig:popgrowth} (left) shows population growth, defined here as the population at the end of the year divided by the population at the beginning of the year, in years from 1960 to 2014 in Africa, Asia, Europe, and North America, and Latin America, total in 112  
countries with more than one million inhabitants in 1950 \citep{NagyEtal2017,DaiEtal2020}. Figure \ref{fig:popgrowth} (right) shows the GDP of every country in the study discounted to the 1960 USD. The discounting was performed according to USD inflation. The data was obtained from the World bank. The missing values of the GDP of a country were extrapolated using the closest known ratio of the GDP of the country and the median GDP in that year. The missing values of GDP were interpolated using linear interpolation of the two closest ratios.

We explore the functional general linear model (GLM) for the population growth across years 1960-2014 with categorical covariate (continent), continuous covariate (GDP), and interactions (GDP with respect to the continent). For this purpose, we applied three tests, and on each of them, we applied the proposed FDR control (IATSE) to obtain all years which are significant for the studied covariate. We follow the graphical functional GLM testing procedure introduced for FWER control in \citet{MrkvickaEtal2021a}.

\begin{figure*}[h]
    \centering
    \includegraphics[width=0.9\textwidth]{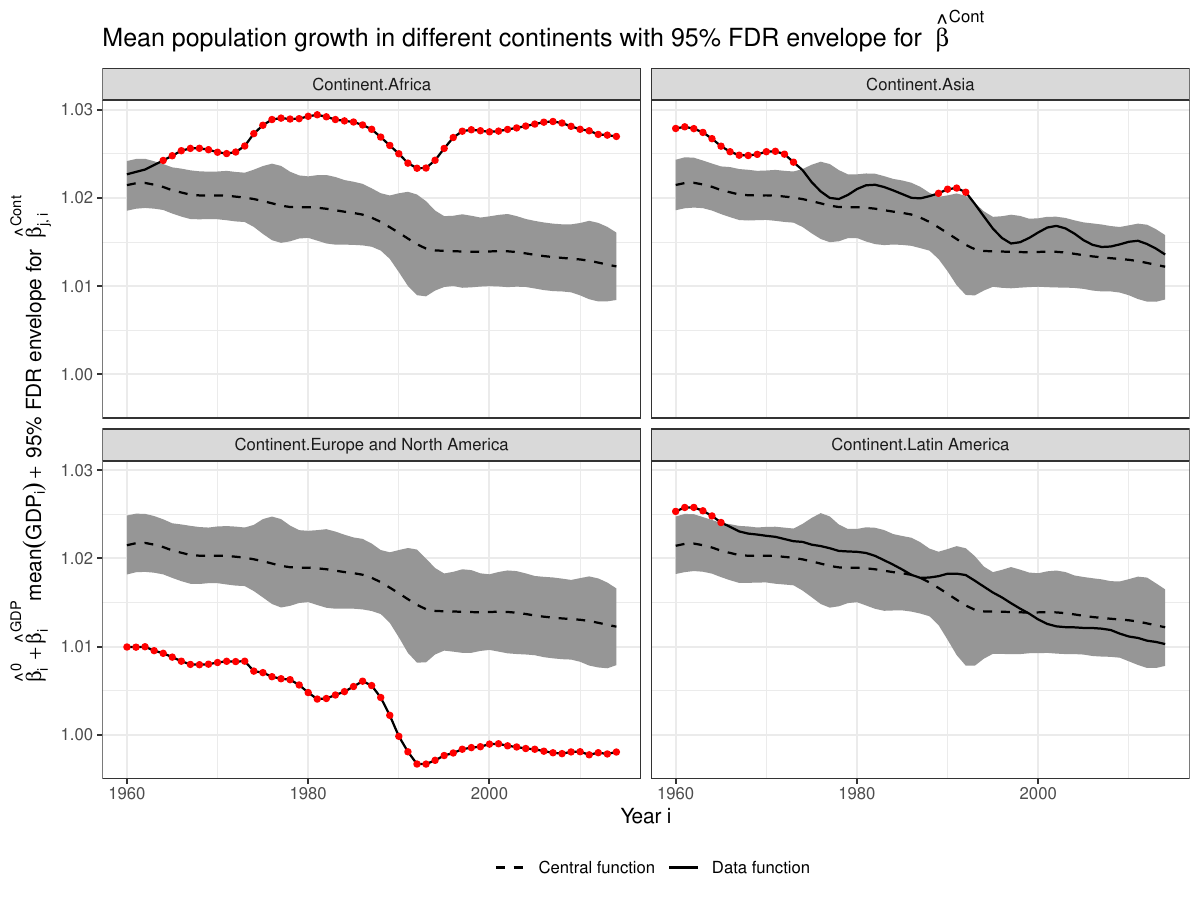}
\caption{\label{fig:popgrowth1}Test for the main effect of the continent, given the GDP (in 1960 USD) as a nuisance factor. The IATSE algorithm was used for FDR control with 5000 permutations. Red color is used to highlight the (significant) years where the data function ($\hat{\beta}_{i}^0 + \hat{\beta}_{i}^\text{GDP}\cdot \text{mean}(\text{GDP}_i) + \hat{\beta}_{j,i}^{\text{Cont}}$, $i=1960,\dots,2014$, $j=1,2,3,4$) goes outside the 95\% FDR envelope. The central function is given by $\hat{\beta}_{i}^0 + \hat{\beta}_{i}^\text{GDP}\cdot \text{mean}(\text{GDP}_i)$.}
\end{figure*}

First, to test the effect of the continent, we assume the main effects model 
$$
\text{population growth} \sim \text{Continent} + \text{GDP}, 
$$
and we used the test statistic 
\begin{eqnarray}\label{TT1}
\nonumber &\Bbeta^{\text{Cont}} = (\underbrace{\beta_{1, 1960}^{\text{Cont}}, \ldots , \beta_{1, 2014}^{\text{Cont}}}_{1:\, \text{Africa}}, 
     \underbrace{\beta_{2, 1960}^{\text{Cont}}, \ldots , \beta_{2, 2014}^{\text{Cont}}}_{2:\, \text{Asia}},\\ & 
     \underbrace{\beta_{3, 1960}^{\text{Cont}}, \ldots , \beta_{3, 2014}^{\text{Cont}}}_{3:\, \text{Europe and North America}}, 
     \underbrace{\beta_{4, 1960}^{\text{Cont}}, \ldots , \beta_{4, 2014}^{\text{Cont}}}_{4:\, \text{Latin America}}),
\end{eqnarray}
where $\beta_{1,i}^{\text{Cont}}, \ldots \beta_{4,i}^{\text{Cont}}$ are the parameters of the univariate GLM model for year $i$, computed with condition $\sum_{j=1}^4 \beta_{j,i}^{\text{Cont}} = 0$. The advantage of this test statistic with respect to the $F$ statistic is that we directly test which continents differ from the mean. 
The permutations were obtained by the standard \citet{FreedmanLane1983} procedure, i.e.\ permuting residuals under the null model ($\text{population growth} \sim \text{GDP}$). The graphical output of the IATSE algorithm is shown in Figure \ref{fig:popgrowth1}. Here, for better interpretation, instead of showing $\Bbeta^{\text{Cont}}$ only, we show the fitted curve for every continent, i.e. the intercept $\beta_{i}^0$ and mean effect of GDP, $\beta_{i}^\text{GDP}\cdot \text{mean}(\text{GDP}_i)$, are added to every $\beta_{j,i}^{\text{Cont}}$. The output also shows the 95\% FDR envelope together with the red points that identify the years and continents, which are significantly different from the fitted mean curve under FDR control. The output shows higher population growth in Africa from 1965 onwards, in Asia for years up to 1974, and in Latin America for years up to 1965. Further, it shows lower population growth for Europe and North America. One could assume that this would be caused by different GDPs in the continents, but GDP is treated here as a nuisance factor. Also, the second test shows only a little importance of the main effect of GDP.

\begin{figure}
    \centering
    \includegraphics[width=0.5\textwidth]{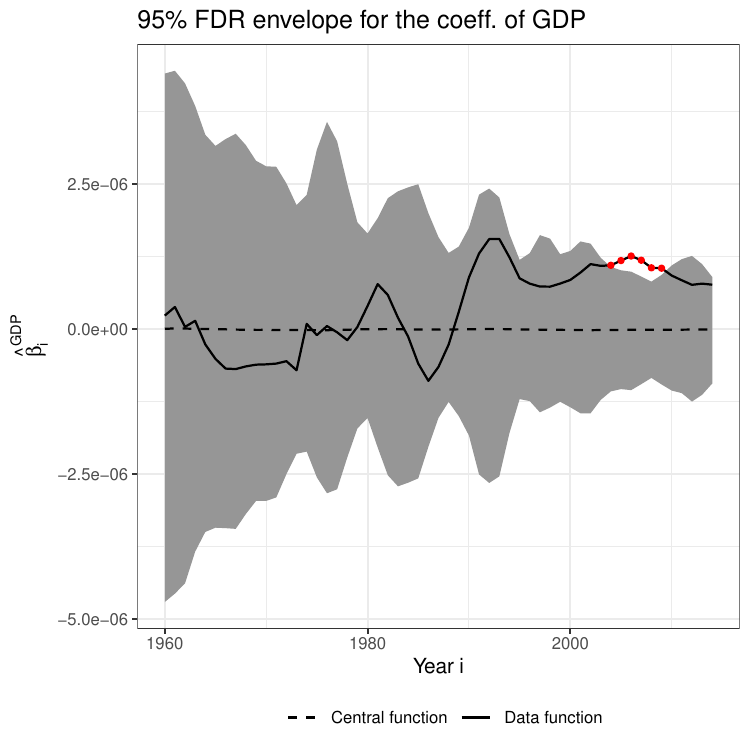}
\caption{\label{fig:popgrowth2}Test for the main effect of GDP (in 1960 USD), given the continent as a nuisance factor. The IATSE algorithm was used for FDR control with 5000 permutations.}
\end{figure}

Second, to test the effect of the GDP, we assumed the same main effects model as above 
and we used simply the test statistic 
\begin{equation}\label{TT2}
\Bbeta^{\text{GDP}} = (\beta_{1960}^{\text{GDP}}, \ldots , \beta_{2014}^{\text{GDP}}).
\end{equation}
The graphical output of the IATSE algorithm is shown in Figure \ref{fig:popgrowth2}. The FDR envelope shows, in this case, rather decreasing variability of the test statistics, which makes no problem to the procedure itself, but it is worth noting for the interpretation. The main effect of GDP is significant only between the years 2004 and 2009.

\begin{figure*}[h]
    \centering
    \includegraphics[width=0.9\textwidth]{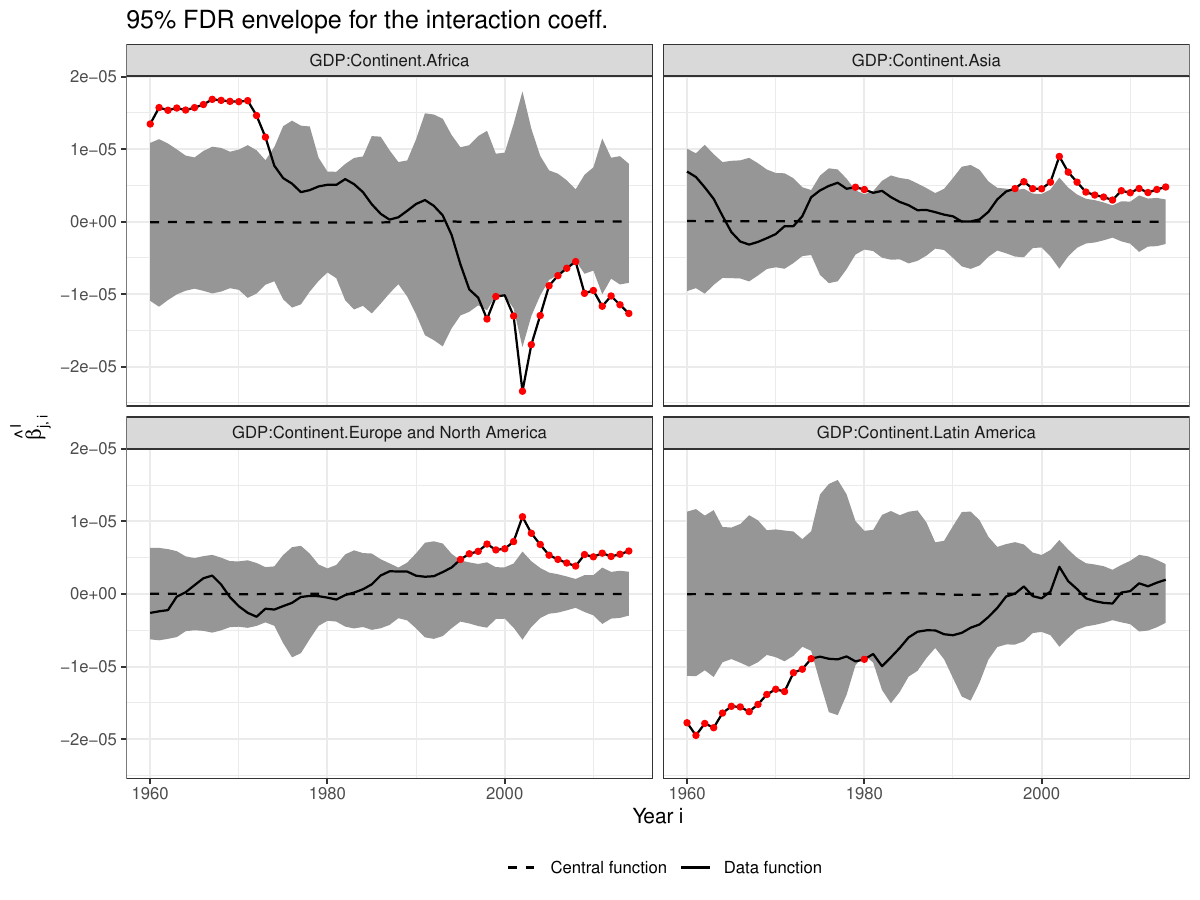}
\caption{\label{fig:popgrowth3}Test for the interaction effect of continent and GDP (in 1960 USD) using the IATSE algorithm for FDR control with 5000 permutations. The null model includes the main effects.}
\end{figure*}

Third, to test the effect of the interactions, we assume the factorial model 
$$
\text{population growth} \sim \text{Continent} * \text{GDP}, 
$$
and we used the test statistic 
\begin{eqnarray}\label{TT3}
\nonumber &\Bbeta^I = (\underbrace{\beta_{1, 1960}^I, \ldots , \beta_{1, 2014}^I}_{1:\, \text{Africa}}, 
     \underbrace{\beta_{2, 1960}^I, \ldots , \beta_{2, 2014}^I}_{2:\, \text{Asia}}, \\ & 
     \underbrace{\beta_{3, 1960}^I, \ldots , \beta_{3, 2014}^I}_{3:\, \text{Europe and North America}}, 
     \underbrace{\beta_{4, 1960}^I, \ldots , \beta_{4, 2014}^I}_{4:\, \text{Latin America}}),
\end{eqnarray}
where $\beta_{1,i}^I, \ldots \beta_{4,i}^I$ are interaction parameters of the univariate GLM model for year $i$, computed with condition $\sum_{j=1}^4 \beta_{j,i}^I = 0.$  
The permutations of residuals from the null model ($\text{population growth} \sim \text{Continent} + \text{GDP}$) was used. The graphical output of the IATSE algorithm is shown in Figure \ref{fig:popgrowth3}. It provides rich interpretability of the interaction component. The population growth of countries in Africa is significantly positively influenced by GDP up to 1974, whereas it is significantly less influenced than the whole world from 1998 onwards. The population growth of Asian countries is more positively influenced by GDP than the whole world from 1996 onwards. The same holds for Europe and North America. On the other hand, the population growth in Latin American countries is negatively influenced by GDP up to 1974. 

We also applied ATS to the test vectors \eqref{TT1}, \eqref{TT2} and \eqref{TT3}. ATS rejected one year less than the FDR envelope in total, and it gave no information about the reasons (direction) of rejection and variability of the test statistic under the null. 
Definitely, the results of FDR envelope presented in Figures \ref{fig:popgrowth1}, \ref{fig:popgrowth2} and \ref{fig:popgrowth3} are much richer and allow for interpreting the results in much deeper detail than the results provided by ATS or other non-graphical FDR algorithm which gives only $q$-values or the rejections as an output.
For comparison, if the $F$-statistic is used instead of the test vector \eqref{TT1} and \eqref{TT3}, we obtain even less information from the test. This $F$-test of the main effect of the continent just rejected the hypotheses for all years, and this test of the interaction effect rejected the hypotheses for the years 1960-1982, 1988, and 1995-2014.

The envelopes of Figures \ref{fig:popgrowth2} and \ref{fig:popgrowth3} are rather fluctuating due to fluctuations in the GDB values \ref{fig:popgrowth}. 
To reduce the fluctuation of these envelopes the standardised beta coefficients can be used instead of beta coefficients in Formulas \ref{TT2} and \ref{TT3}. Since the standardisation has no effect on the significance of these envelopes, we prefer the simpler versions.  

\section{Local spatial correlation}\label{sec:localcor}

In spatial statistics, it is a relevant question to ask in what regions two random fields are correlated. This question was studied nonparametrically in \citet{Viladomat2014}, where the pointwise local $p$-values were obtained. \citet{Viladomat2014} did not perform any adjustment for multiple testing. The aim of this section is to show the possibility of FDR control in a 2D example with the use of our proposed FDR envelopes.

Assume that we observe two random fields $\Phi$ and $\Psi$ in the observation window $W\subset \mathbb R^2$. 
Let $X$ be the set of sampling locations where the values of $\Phi$ are observed. $X$ can be either random or non-random, depending on the design of the experiment. We denote by $\Phi(X)$ the vector of values of $\Phi$ observed at sampling points $X$, and similarly for $\Psi(X)$.

The functional test statistic describing the local correlation at a location $u\in W$ can be calculated as 
\begin{eqnarray}
    \nonumber \hat r_b(u) &= \frac{\sum_{x\in X}w_b(u-x)(\Phi(x)-\overline{\Phi(x)})(\Psi(x)-\overline{\Psi(x)})}{\sqrt{\sum_{x\in X}w_b(u-x)(\Phi(x)-\overline{\Phi(x)})^2}}\cdot\\
    &\frac{1}{\sqrt{\sum_{x\in X}w_b(u-x)(\Psi(x)-\overline{\Psi(x)})^2}},
\end{eqnarray}
where $w_b(s)$ is a kernel with bandwidth $b$ (we have used a Gaussian kernel with bandwidth equal to its standard deviation), 
$$
\overline{\Phi(x)}=\frac{\sum_{x\in X}w_b(u-x)(\Phi(x))}{\sum_{x\in X}w_b(u-x)}
$$
and 
$$
\overline{\Psi(x)}=\frac{\sum_{x\in X}w_b(u-x)(\Psi(x))}{\sum_{x\in X}w_b(u-x)}. 
$$
The asymptotic behavior of $\hat r_b$ is unknown; therefore, resampling under the null model of independence has to be performed. \citet{Viladomat2014} proposed to obtain resamples $\Psi^*$ by smoothing the permuted values $\Psi(x)$ in order to obtain a smoothed empirical variogram that matches its counterpart for $\Psi$. \citet{MrkvickaEtal2021b} proposed the random shift method with a variance correction for the test statistic instead. They showed that random shifts are more powerful in assessing the global correlation. However, for assessing the local correlation, the number of points used to calculate $\hat r_b(u)$ would have to be used in the mentioned variance correction. Since for local correlation, this issue would have to be investigated, we use the \citet{Viladomat2014} proposal instead. 

\begin{figure*}[h]
    \centering
    \includegraphics[width=0.9\textwidth]{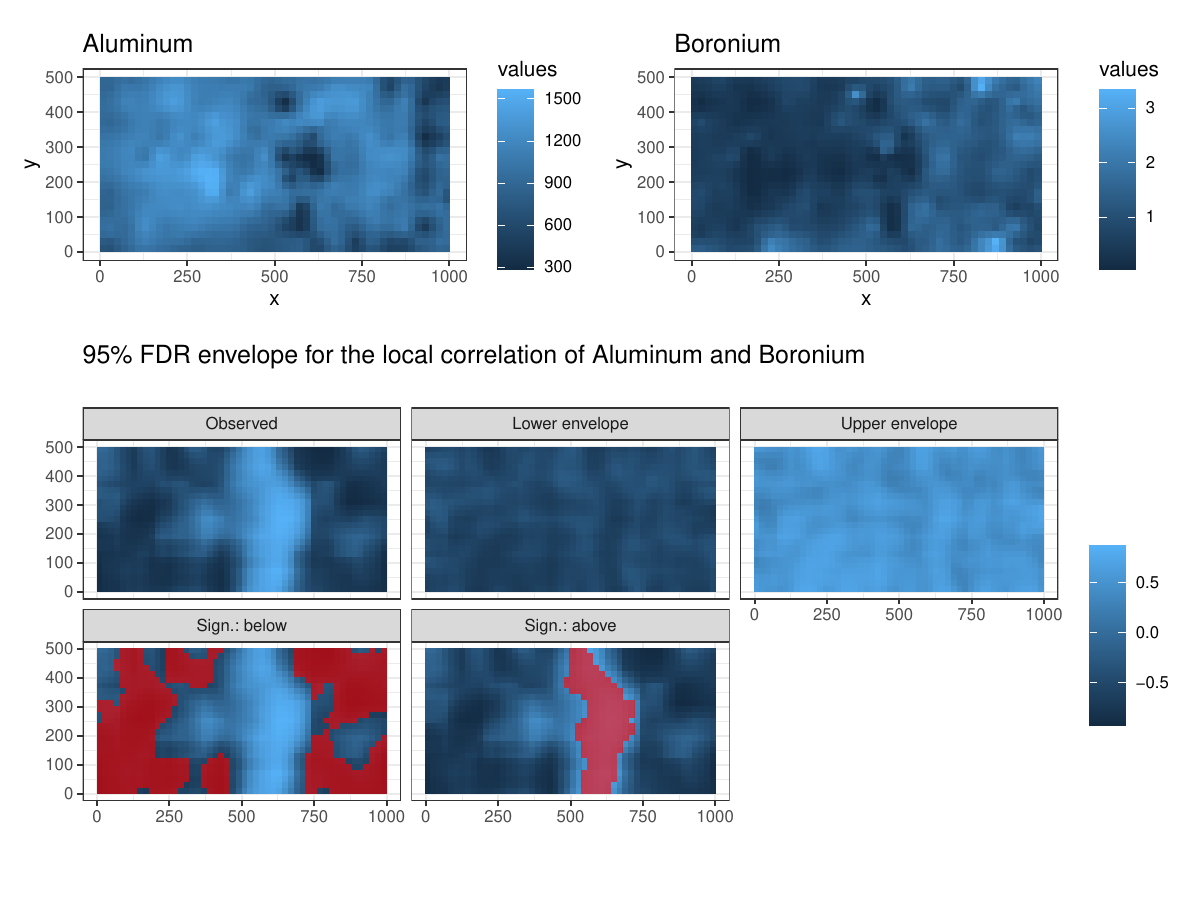}
\caption{\label{fig:localcorr}Top: Values of Aluminium and Boronium in the tropical forest plot of area $1000\text{ m} \times 500\text{ m}$. Bottom: The test of local correlations between Aluminium and Boronium under FDR control with the IATSE algorithm: observed local correlations, the lower and upper boundary of the FDR envelope, and regions (red) where the local correlation is significantly below or above the FDR envelope.}
\end{figure*}

For the illustration we use the covariates observed in the tropical forest plot of area $1000\text{ m} \times 500\text{ m}$ in Barro Colorado Island \citep{Hubbell2005,Condit1998,Hubbell1999}.
This data set is accompanied with many covariates but, for the purpose of illustration, we study only the dependence of the first two covariates, i.e., Aluminium (Al) and Boronium (B) observed on a $50 \times 25$ grid with equally spaced observations (see Figure~\ref{fig:localcorr}, top). 
The global correlation (a single Pearson's correlation coefficient) between the two covariates is approximately $-0.41$, which is significant according to the \citet{Viladomat2014} global test of independence ($p\approx 0.031$).
We then applied the described local independence test (with $b=100$) to find all regions with significant local correlations. The results in Figure \ref{fig:localcorr} (bottom) show that there are significant positive local correlations in the middle of the image, whereas on the left and right side of the image, there are significant negative local correlations. 

\section{Discussion}\label{sec:discussion}

The aim of this work was to find a good method for determining the FDR envelope. 
For this purpose, we defined two alternatives, namely ATSE  and IATSE.
These algorithms are based on the idea of \citet{StoreyTibshirani2001}, but the estimation methods of $\pi_0$ introduced in \citet{Storey2002} and \citet{StoreyTibshirani2003} were found to be unsatisfactory for highly dependent test statistics which arise in functional data analysis. Therefore, the proposed algorithms are supplied with different $\pi_0$ estimates. 
Based on our study, these two methods perform as well as the ATS algorithm, which we have found to be the best from the literature, both in terms of being close to the nominal FDR level and the power. 
The simulation study also suggests that the IATSE algorithm is a little bit more powerful than the ATS algorithm, especially for lower numbers of replicates. It can be seen as an advantage when the simulation of the replicates is time-consuming as it can be, e.g., in spatial statistics. 
We also investigated resampling methods capturing the dependence between tests; unfortunately, the only method applicable in the required wide generality turned out to be rather liberal if the method was made to be adaptive, i.e., when the estimate of $\pi_0$ was considered in the algorithm.

Since our proposed methods construct the rejection region from resamples and their pointwise ranks, these methods are robust with respect to changes in the distribution of the test statistic along its domain. This is not the case for the methods using the parametric $p$-values, which require the homogeneity of the test statistic across the domain. The same can be said for the homogeneity of the correlation structure of the test statistic along its domain. 

The population growth example (Section \ref{sec:population}) shows the interpretation advantages of FDR envelopes with respect to pure $q$-values, which can be obtained from other FDR methods without graphical interpretation. 
The local independence test (Section \ref{sec:localcor}) shows the possibility of applying the FDR envelope in the 2D case in spatial statistics.
The greatest advantages of our proposed methods are the graphical interpretation, its generality, and its easy application via the R package GET \citep{MyllymakiMrkvicka2020}.  
The generality of our methods includes that the methods can be used with any test statistic without knowing its distribution, the distribution of the test statistic can vary along its domain, and the autocorrelation of the test statistic can also vary. 

The method of \citet{StoreyTibshirani2003} was promoted in the context of permutation tests in genetics studies, where the dependence between the test statistics is not as strong as in functional data analysis. For this method, 
\citet{JiaoZhang2008} and \citet{XieEtAl2005} proposed to estimate $\mathbb E \RR^0(\Gamma)$ only from those hypotheses which are true null for the resampling approach. Obviously, it is impossible to know the true null hypotheses, and therefore they used only hypotheses which are not rejected for the data. The reason to avoid false null hypotheses was the increase of variance for false null with respect to true null. Since our rejection region, $\Gamma_\gamma$, is not a constant number, as it is usually assumed, but it is constructed from resampling of the whole test vector under the complete null hypothesis, $\Gamma_\gamma$ is adapted to this increased variability (see, e.g., Figure \ref{fig:popgrowth2}). This means that the whole problem of the \citet{StoreyTibshirani2001} procedure discussed in \citet{JiaoZhang2008} and \citet{XieEtAl2005} is overcome by our construction of $\Gamma_\gamma$. 

Our proposed methods are defined under the assumption of simple null hypothesis testing. Additionally, we introduced an algorithm based on these methods and the quadratic Monte Carlo scheme, which makes the FDR envelopes valid also for composite null hypotheses (see Appendix \ref{sec:composite}). The performance of this algorithm with respect to neglecting the compositeness of the null hypothesis will be studied in the future.

\section*{Acknowledgements}
T.\,M.\ was financially supported by the Grant Agency of the Czech Republic (project number 19-04412S), 
and M.\,M.\ by the Academy of Finland (project numbers 295100, 306875, 327211). 
The BCI forest dynamics research project was founded by S.P. Hubbell and R.B. Foster and is now managed
by R. Condit, S. Lao, and R. Perez under the Center for Tropical Forest Science and the Smithsonian
Tropical Research in Panama. Numerous organizations have provided funding, principally the U.S.
National Science Foundation and hundreds of field workers have contributed. The Barro Colorado
Island soils data set was collected and analyzed by J. Dalling, R. John, K. Harms, R. Stallard, and J.
Yavitt, with support from National Science Foundation, grants DEB021104, DEB021115, DEB0212284,
DEB0212818 and OISE 0314581, Smithsonian Tropical Research Institute and Center for Tropical
Forest Science. 
The authors thank Mikko Kuronen (Luke) for his remarks on an earlier version of the manuscript.

\begin{appendices}

\section{Composite null hypothesis}\label{sec:composite}

When the null hypothesis is not simple, but some nuisance parameters are estimated, the estimate (5) of the main document 
will be biased, and thus the estimate of FDR will also be biased. In such case, an adjusted estimate of $\mathbb E \rho^0(\Gamma_\gamma)$ can be performed in a similar manner as for Monte Carlo methods controlling FWER \citep{DaoGenton2014, BaddeleyEtal2017}. 

Assume that the null hypothesis has a parameter $\theta$, which needs to be estimated in order to obtain simulations from the null hypothesis, e.g., for a goodness-of-fit test with a functional test statistic. To estimate $\mathbb E \rho^0(\Gamma_\gamma)$ a theoretical density of $\hat\theta$ estimator can be used: 
$$\mathbb E \rho^0(\Gamma_\gamma)=\int \mathbb E (\rho^0(\Gamma_\gamma)\mid\hat\theta) p(\hat\theta) d\hat\theta.$$
For $p(\hat\theta)$, an asymptotic distribution of $\hat\theta$ or, more preferably, a two-stage Monte Carlo scheme \citep{BaddeleyEtal2017} can be used.

\begin{algorithm}\label{alg:TWMC}
(Quadratic) Two-stage Monte Carlo estimate of $\mathbb E \rho^0(\Gamma_\gamma)$:
\begin{enumerate}
\item Estimate $\hat\theta$.

\item Simulate $s_1$ simulations from the fitted model with $\hat\theta$. 

\item Estimate $\hat\theta_i, i=1, \ldots , s_1,$ for every simulation.

\item Simulate $s_2$ simulations from $\hat\theta_i$ for every $i$, $i=1, \ldots , s_1$.

\item Compute $\mathbb E (\rho^0(\Gamma_\gamma)\mid\hat\theta_i)$ for every $i$ via 
$$\mathbb E \RR^0(\Gamma_\gamma\mid\hat\theta_i) \approx \frac{1}{s_2}\sum_{j=1}^{s_2} \RR^0_j(\Gamma_\gamma),$$ where $\RR^0_j(\Gamma_\gamma)$ is the number of rejections produced by $\Gamma_\gamma$ on the $j$-th simulation from $\hat\theta_i$.

\item Estimate $\mathbb E \rho^0(\Gamma_\gamma)$ by average of $\mathbb E (\rho^0(\Gamma_\gamma)\mid\hat\theta_i)$. 
\end{enumerate}
\end{algorithm}

Thus, this algorithm requires $s$ simulations for constructing $\Gamma_\gamma$ and $s_1\cdot s_2$ simulations for computing the expectation $\mathbb E \rho^0(\Gamma_\gamma)$. Since the quadratic Monte Carlo method is used to approximate the expectation, the numbers $s_1$ and  $s_2$ can be small, and the whole algorithm will not be too computational. 

\section{Incorporating dependence of the test statistics in the FDR estimation}\label{sec:dep}

\citet{YekutieliBenjamini1999} provided a resampling method under the assumption of subset pivotality and independence of null and alternative hypotheses in order to handle the dependence between test statistics. \citet{SchwartzmanLin2011} showed that correlation may increase the bias and variance of the FDR estimators substantially in comparison to the independent case. 
In order to solve this issue, some authors modeled the correlation, e.g., \ \citet{SunCai2009} modeled the correlation by a hidden Markov model and, on the other hand, \citet{RomanoEtAl2008} modeled the correlation via bootstrapping of the original data. 
Since we do not want to rely on model assumptions, only the method of \citet{YekutieliBenjamini1999} is usable. Inspired by it, we also investigate the adaptive FDR versions of \citet{YekutieliBenjamini1999} estimators for correlated test statistics. 

Let us first describe the original \citet{YekutieliBenjamini1999} procedure which is based on $p$-values, in our case resampling $p$-values, and which consists of one step only. The FDR for a given threshold $p$ (defining the rejection region $\Gamma$) is estimated by
\begin{equation}\label{dependence}
\mathbb E \left[ \frac{V(p)}{\RR(p)}\right] \approx   \mathbb E Q^*(p) ,
\end{equation}
where
\begin{eqnarray*}
Q^*(p) &=& 0 \quad  \text{if} \quad \RR^0(p)=0,  \\
Q^*(p) &=& 1 \quad  \text{if} \quad \RR^0(p)>0 \ \text{and} \ \rho(p) - \rho_u^0(p) < m p, \\
Q^*(p) &=& \frac{ \RR^0(p)}{\RR^0(p)-m p+\RR(p)} \quad  \text{if} \quad \RR^0(p)>0 \ \text{and} \\ && \rho(p) - \rho_u^0(p) \geq m p. 
\end{eqnarray*}
Here $\rho(p)$ is the number of rejections with the threshold $p$, i.e. $\mid\{k: p_k \leq p \}\mid$, $\RR^0(p)$ is the number of rejections with the threshold $p$ of resampled data, i.e. $\mid\{k:p_k^* \leq p \}\mid$. The resampled data are obtained here under the complete null hypothesis, similarly as our envelopes $\mathcal{E}$. Finally, $V(p)$ is the number of false rejections with the threshold $p$. 

\citet{YekutieliBenjamini1999} also defined a more conservative estimate with 
\begin{eqnarray*}
Q^*_u(p) &=& 0 \quad  \text{if} \quad \RR^0(p)=0,  \\
Q^*_u(p) &=& 1 \quad  \text{if} \quad \RR^0(p)>0 \ \text{and} \ \rho(p) - \rho_u^0(p) < 0, \\
Q^*_u(p) &=& \frac{ \RR^0(p)}{\RR^0(p)-\rho_u^0(p)+\RR(p)} \quad  \text{if} \quad \RR^0(p)>0 \ \text{and} \\ && \rho(p) - \rho_u^0(p) > 0.
\end{eqnarray*}
Here $\rho_u^0(p)$ is defined as 1-$\beta$ quantile of $\RR^0(p)$, where \citet{YekutieliBenjamini1999} used $\beta=0.05$.

Since we do not assume independence of null and alternative hypotheses, these estimators are theoretically not valid. Indeed, resampling approximates the marginal distribution of $V(p)$, but not the conditional distribution of $V(p)$ given
$\RR(p)=r$. We anyway modified these algorithms to our setting with rejection region $\Gamma_\gamma$ and replaced steps 1. and 3. of the ATSE and IATSE algorithms 
with it. This defines four new adaptive FDR estimators, which should handle the dependence between the test statistics. We do not provide a rigorous definition of these algorithms since they are found to be liberal in the simulation study. Note here only that we performed two independent sets of resamples here, first for obtaining $\Gamma_\gamma$ and second for estimating $Q^*(\Gamma_\gamma)$ and $Q^*_u(\Gamma_\gamma)$.

\section{Results of the simulation study}\label{sec:simstudy}

Here we complete the results of the simulation study of Section \ref{sec:SS} of the main document.
For the details of the simulation study setup, we refer to Section \ref{sec:SS} of the main document. 
In particular, Table \ref{tab:simstudy_methods} there explains the methods that were compared in the simulation study (shown in all plots on the y-axis), and Table \ref{tab:simstudy_means} explains the considered models. Figure \ref{fig:simstudy_means} visualises the mean functions of Table \ref{tab:simstudy_means}.
In all figures below, $s$ denotes the number of resamples, and $\phi_Z$ is the correlation parameter of the error term.

\begin{figure*}
    \centering
    \includegraphics[width=\textwidth]{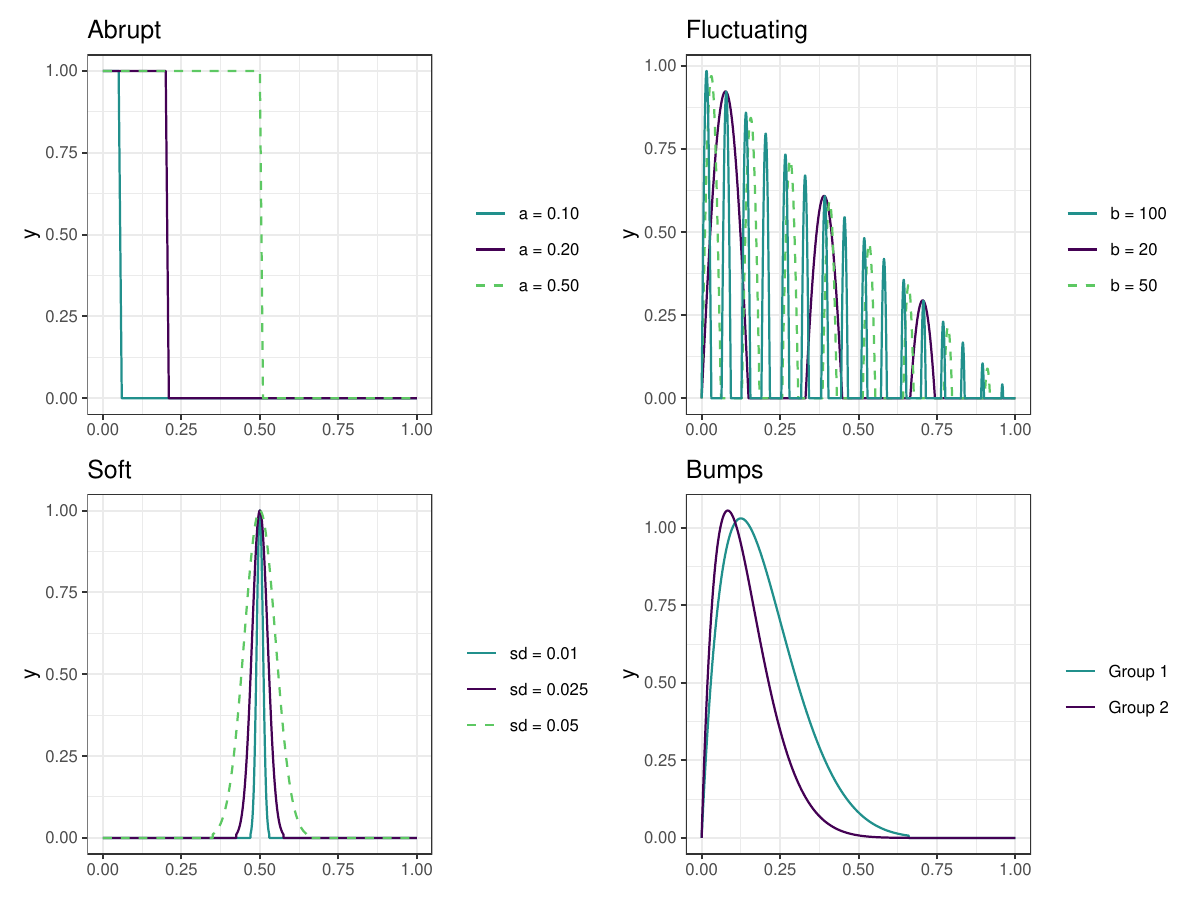}
    \caption{Mean functions in the simulation study for the second group for 'Abrupt', 'Fluctuating' and 'Soft' with different parameter values and for both groups for 'Bumps' (see Table 3).}
    \label{fig:simstudy_means}
\end{figure*}

\begin{figure*}
    \centering
\includegraphics[width=\textwidth]{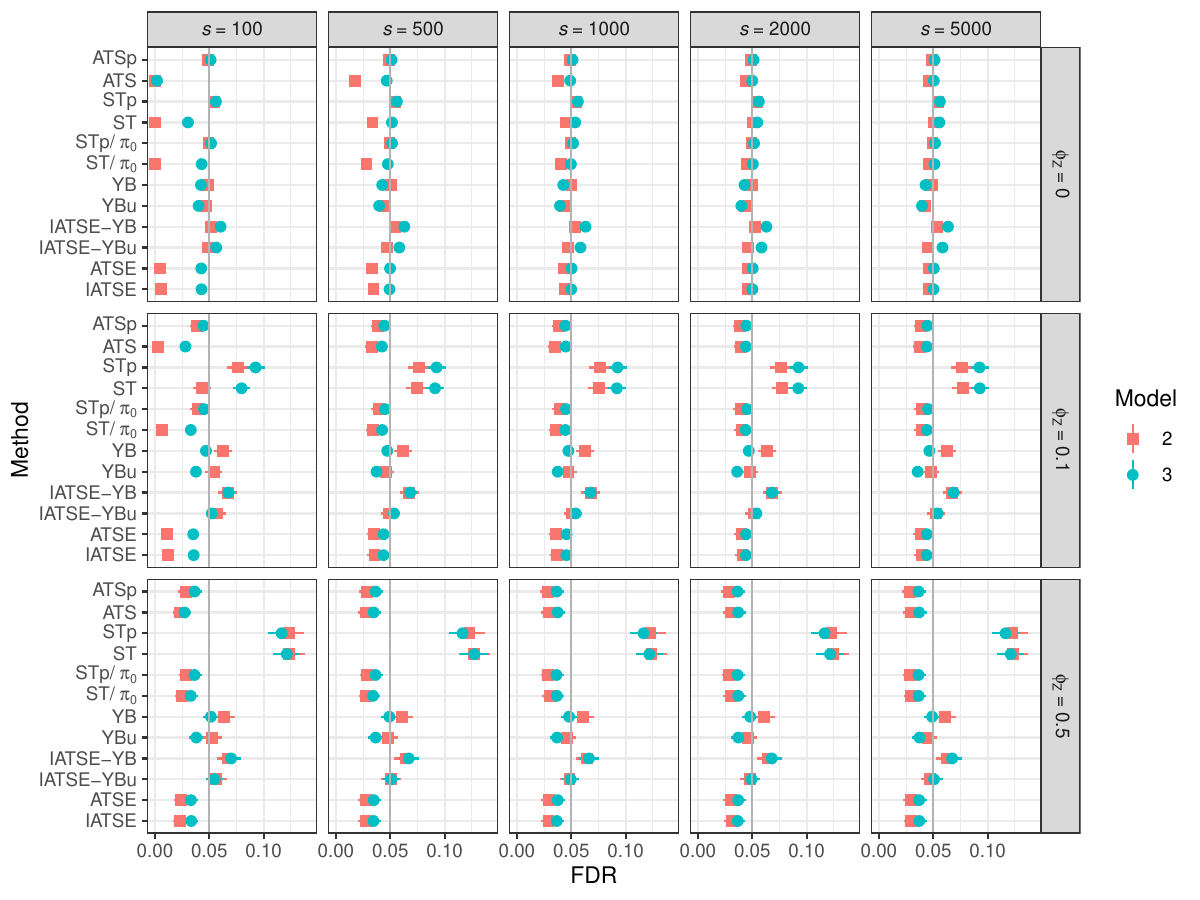}
    \caption{\label{fig:res_part1_model23}The mean FDR ($\pm 2\cdot \sigma_{\text{FDR}}$ given by bars around dots) of the different methods for Model 2 and 3 with Gaussian error and $m=200$.}
\end{figure*}

\begin{figure*}
    \centering
\includegraphics[width=\textwidth]{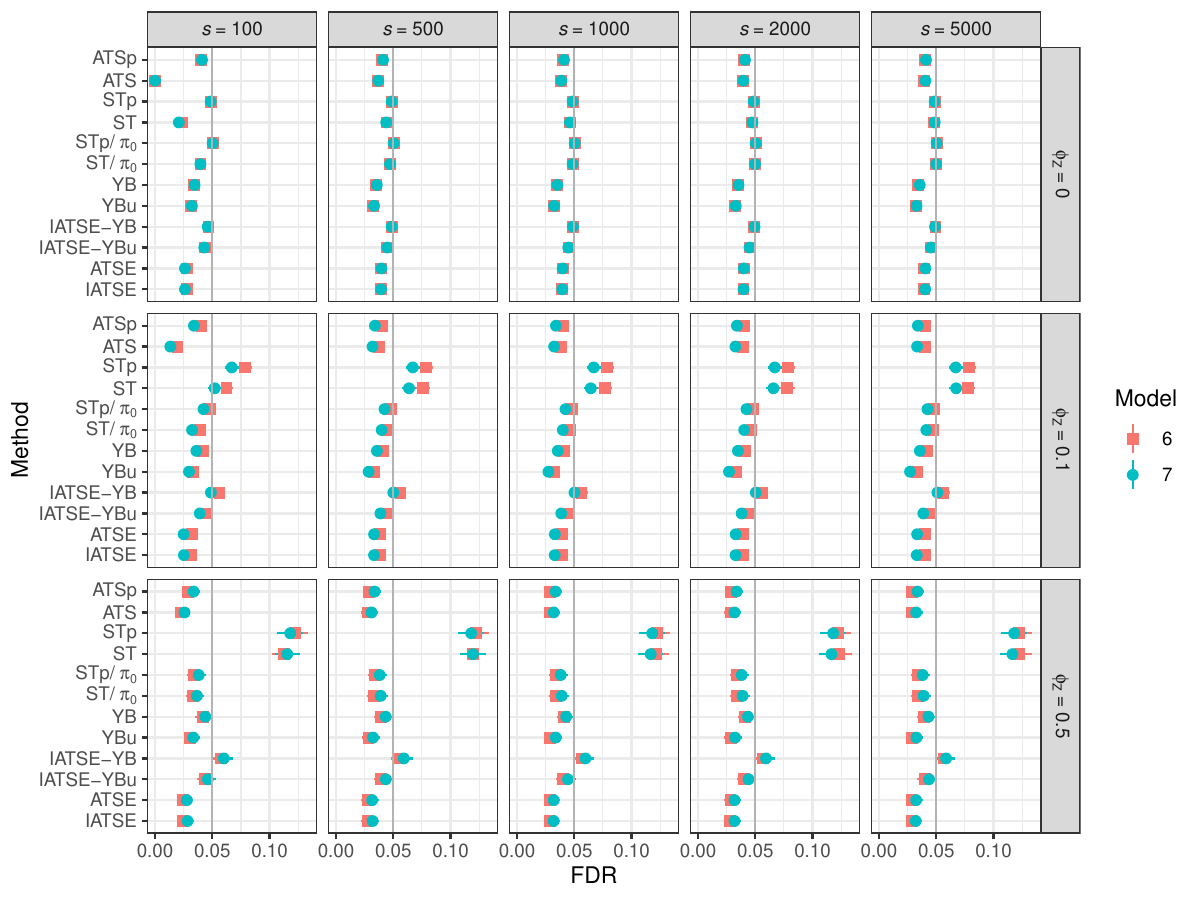}
    \caption{\label{fig:res_part1_model67}The mean FDR ($\pm 2\cdot \sigma_{\text{FDR}}$ given by bars around dots) of the different methods for Model 6 and 7 with Gaussian error and $m=200$.}
\end{figure*}

\begin{figure*}
    \centering
\includegraphics[width=\textwidth]{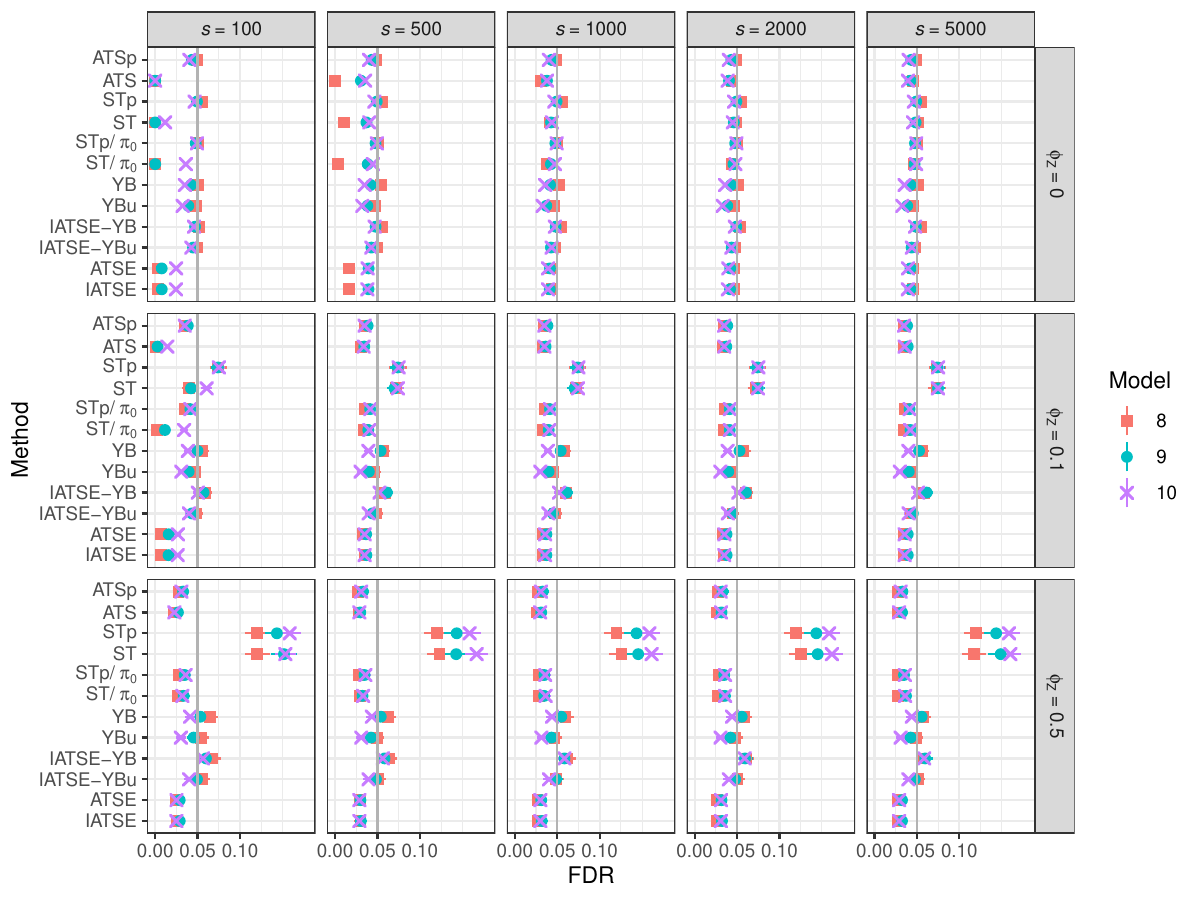}
    \caption{\label{fig:res_part1_model8910}The mean FDR ($\pm 2\cdot \sigma_{\text{FDR}}$ given by bars around dots) of the different methods for Model 8, 9 and 10 with Gaussian error and $m=200$.}
\end{figure*}

\begin{figure*}
    \centering
    \includegraphics[width=\textwidth]{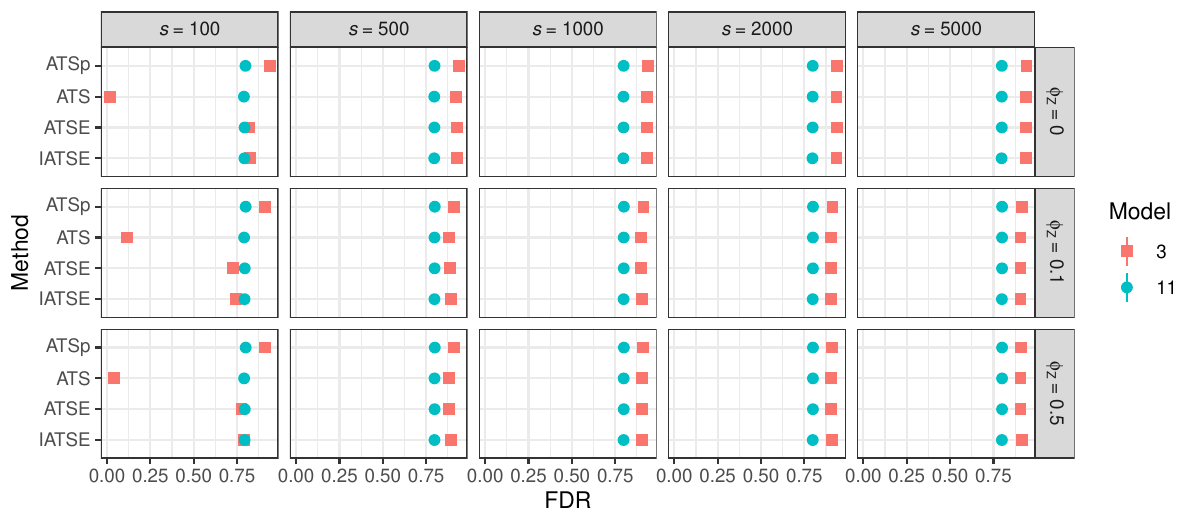}
    \caption{\label{fig:res_part1_power_model311}The mean power ($\pm 2\cdot \sigma_{\text{Power}}$ given by bars around dots) of the four methods for Model 3 and 11 with Gaussian error and $m=200$.}
\end{figure*}

\begin{figure*}
    \centering
    \includegraphics[width=\textwidth]{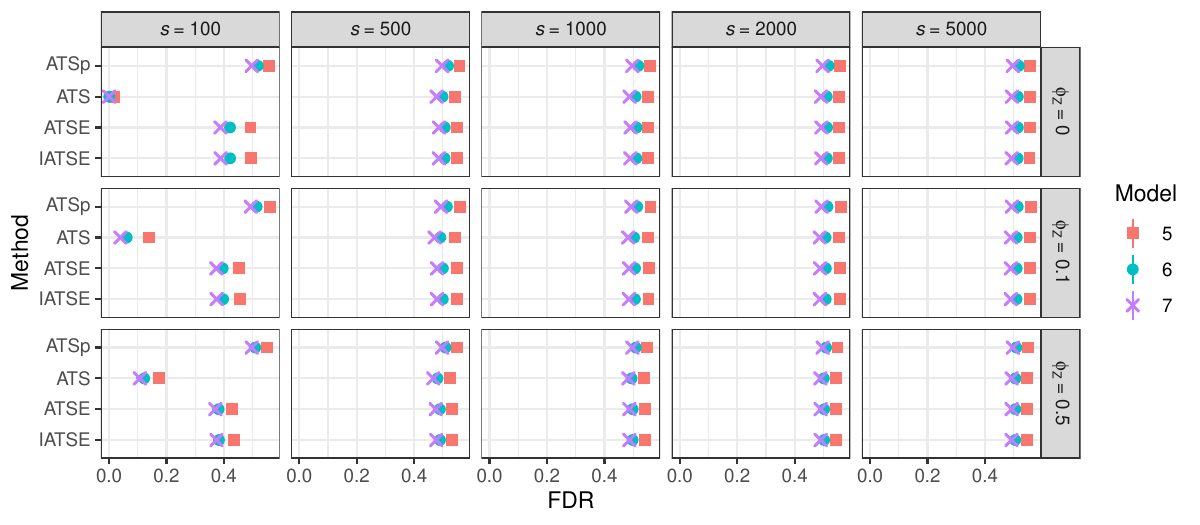}
    \caption{\label{fig:res_part1_power_model67}The mean power ($\pm 2\cdot \sigma_{\text{Power}}$ given by bars around dots) of the four methods for Model 5, 6 and 7 with Gaussian error and $m=200$.}
\end{figure*}

\begin{figure*}
    \centering
    \includegraphics[width=\textwidth]{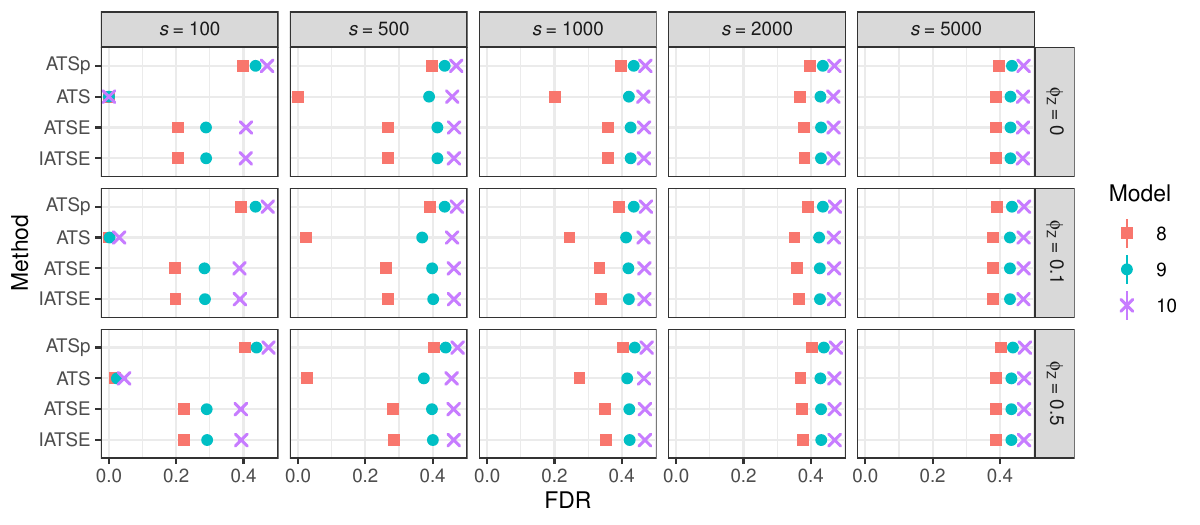}
    \caption{\label{fig:res_part1_power_model8910}The mean power ($\pm 2\cdot \sigma_{\text{Power}}$ given by bars around dots) of the four methods for Model 8, 9 and 10 with Gaussian error and $m=200$.}
\end{figure*}

\begin{figure*}
    \centering
\includegraphics[width=\textwidth]{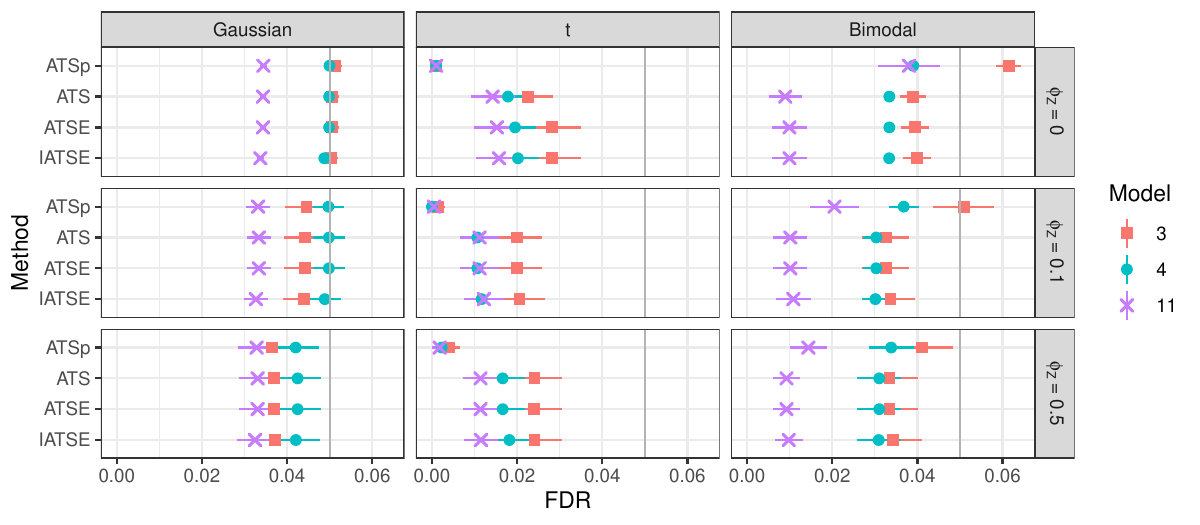}
    \caption{\label{fig:res_part1_errors_model3411}The mean FDR ($\pm 2\cdot \sigma_{\text{FDR}}$ given by bars around dots) of the four methods for Model 3, 4 and 11 with Gaussian error, heavy tailed error ($t$) and bimodal error for $s=5000$ and $m=200$.}
\end{figure*}

\begin{figure*}
    \centering
\includegraphics[width=\textwidth]{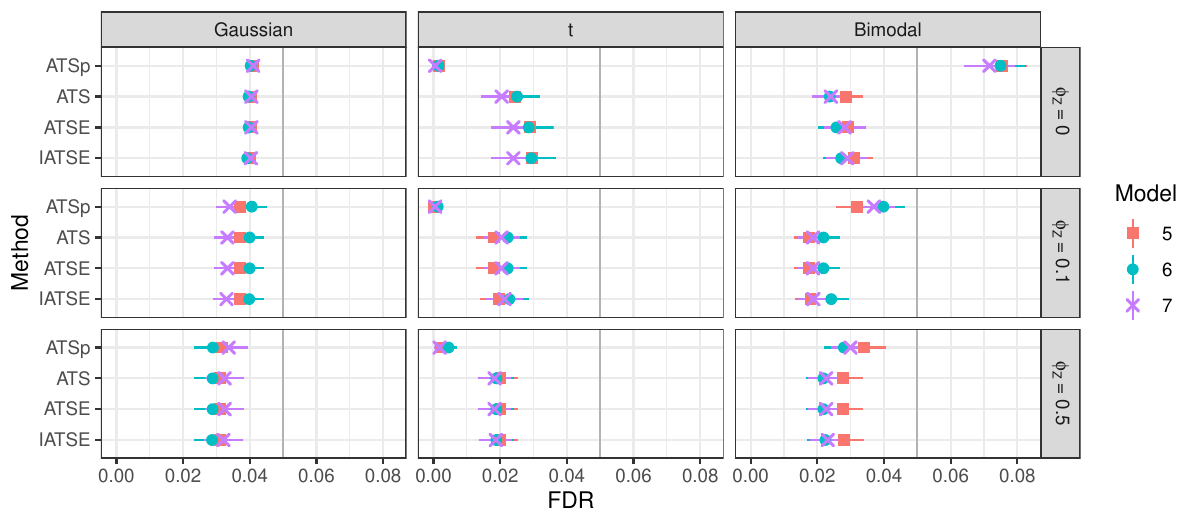}
    \caption{\label{fig:res_part1_errors_model567}The mean FDR ($\pm 2\cdot \sigma_{\text{FDR}}$ given by bars around dots) of the four methods for Model 5, 6 and 7 with Gaussian error, heavy tailed error ($t$) and bimodal error for $s=5000$ and $m=200$.}
\end{figure*}

\begin{figure*}
    \centering
\includegraphics[width=\textwidth]{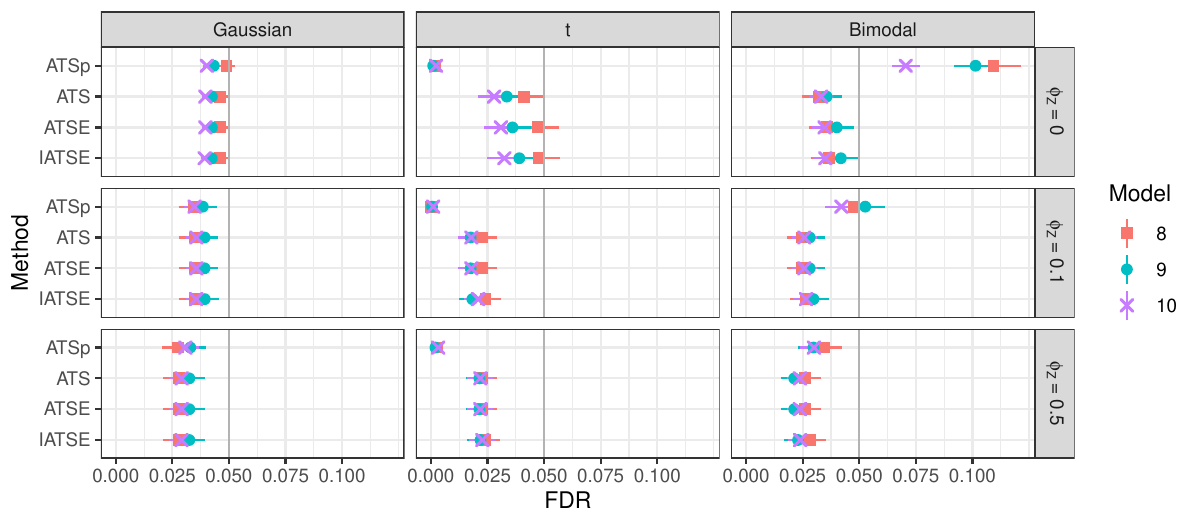}
    \caption{\label{fig:res_part1_errors_model8910}The mean FDR ($\pm 2\cdot \sigma_{\text{FDR}}$ given by bars around dots) of the four methods for Model 8, 9 and 10 with Gaussian error, heavy tailed error ($t$) and bimodal error for $s=5000$ and $m=200$.}
\end{figure*}

Figures \ref{fig:res_part1_model23}, \ref{fig:res_part1_model67} and \ref{fig:res_part1_model8910} present the mean FDR for the case of Gaussian error for the models ((2, 3), (6, 7) and (8, 9, 10), respectively) that were not shown in the main document.
Figures \ref{fig:res_part1_power_model311}, \ref{fig:res_part1_power_model67} and \ref{fig:res_part1_power_model8910} present the mean power for the case of Gaussian error for the models ((3, 11), (6, 7), and (8, 9, 10), respectively) that were not shown in the main document. In these and the following figures, the results are only shown for the four methods (ATS, ATSp, ATSE, and IATSE) that had reasonable FDR levels. 
Finally, Figures \ref{fig:res_part1_errors_model3411}, \ref{fig:res_part1_errors_model567} and \ref{fig:res_part1_errors_model8910} show the comparison of the performance of the methods with respect to the different error structures for the rest of the models not shown in the main document.





\end{appendices}


\end{document}